\documentclass[11pt, a4paper]{article}
\usepackage{array}
\usepackage{graphicx}
\usepackage{amsmath, amssymb, amsthm}
\usepackage{hyperref}
\usepackage{amsfonts}
\usepackage{amssymb}
\usepackage{amsthm}
\usepackage{amsmath}
\usepackage{multirow}
\usepackage{color}
\usepackage{cite}
\usepackage{float}
\usepackage{geometry}
\usepackage{authblk}
\usepackage{hyperref}
\usepackage{cleveref}
\usepackage{amsmath}   
\usepackage{amssymb}   
\usepackage{siunitx}   

\newcommand{\beq}{\begin{equation}}
	\newcommand{\eeq}{\end{equation}}
\newcommand{\bea}{\begin{eqnarray}}
	\newcommand{\eea}{\end{eqnarray}}

\newcommand{\gsim}{\lower.7ex\hbox{$\;\stackrel{\textstyle>}{\sim}\;$}}
\newcommand{\lsim}{\lower.7ex\hbox{$\;\stackrel{\textstyle<}{\sim}\;$}}
\newcommand{\be}{\begin{equation}}
	\newcommand{\ee}{\end{equation}}
\newcommand{\ba}{\begin{eqnarray}}
	\newcommand{\ea}{\end{eqnarray}}

\usepackage{caption}

\usepackage{cancel}

\title{Einstein-Gauss-Bonnet gravity corrections in Dark Dimension scenario}
\author{George K. Leontaris\thanks{Email: \texttt{leonta@uoi.gr}}, \;George Prampromis\thanks{Email: \texttt{g.prampromis@uoi.gr}}\;}

\affil{${}^*$Physics Department, University of Ioannina, 45110, Ioannina, Greece}

\begin{document}
	
	\maketitle

\begin{abstract}
We investigate the thermodynamics, dynamical stability, and potential
phenomenological implications of Einstein--Gauss--Bonnet (EGB) black holes
in Dark Dimension scenarios with one or two micron-sized extra dimensions.
For the asymptotically flat, neutral Boulware--Deser solution with positive
Gauss--Bonnet coupling $\alpha$, we derive the mass, Hawking temperature,
Wald entropy, and heat capacity in arbitrary spacetime dimension $D\geq5$.
A positive-heat-capacity branch occurs only in $D=5$, where the temperature
reaches $T_{\max}=1/(8\pi\sqrt{\alpha})$ at $r_h=2\sqrt{\alpha}$ and
vanishes in the formal zero-radius limit at finite mass
$M_{\rm lim}=3\pi\alpha/(4G_5)$. For
$G_5=(8\pi M_*^3)^{-1}$, $\alpha=M_*^{-2}$, and
$M_*\sim10^{10}\,\mathrm{GeV}$, this gives
$M_{\rm lim}\simeq1.06\times10^{-12}\,\mathrm{g}$.
However, the entire positive-heat-capacity branch lies within the
scalar-type gravitational-instability region of classical EGB gravity.
We examine the relation between these instability bounds and the
ultraviolet cutoff of a string effective description, emphasizing that
excluding unresolved perturbations does not establish the stability of
the background or the existence of a surviving remnant.
Assuming such remnants exist, we discuss their conditional cosmological
abundance, including extended primordial-black-hole mass distributions,
and possible gravitational-wave and evaporation signatures. Quantitative
predictions depend on the formation history and ultraviolet completion.
In $D=6$, the stationary solution has negative heat capacity at all radii
and no analogous zero-temperature endpoint. Within a bulk blackbody
emission model, we derive the leading large-mass lifetime correction and
the formal small-radius scaling $\tau\propto M_0^3$, compared with
$\tau_{\rm GR}\propto M_0^{5/3}$.
For $M_*\sim10\,\mathrm{TeV}$, we also examine geometric production and
decay benchmarks relevant to future colliders. Their interpretation is
limited by tensor-type gravitational instability and, for the assumed
coupling, sub-string-scale horizon radii. The resulting distinction
between stationary EGB predictions and controlled effective-theory
observables is essential when assessing remnant dark matter and collider
phenomenology.
\end{abstract}

	\section{Introduction}
    \label{sec:EGB_Introduction}
	The quest for a consistent quantum theory of gravity remains one of the major open problems  in modern theoretical physics.    
	General Relativity (GR) provides a remarkably successful description of gravitational phenomena over a wide range of scales, but at high  energies, where classical space time singularities arise,  it is generally regarded as an  effective theory that may  require modification or  ultraviolet (UV) completion.      
   An appealing UV complete framework is provided by string theory  where the classical Einstein-Hilbert action can be naturally extended to include  higher-derivative curvature corrections at short distances.  
Among these extensions, Lanczos-Lovelock\cite{Lanczos:1938sf,Lovelock:1971yv} gravity stands out as a mathematically unique framework. In five or more dimensions, its leading order correction is the four-derivative Gauss-Bonnet invariant. A key property of the resulting Einstein–Gauss–Bonnet (EGB) theory is that its equations of motion remain second order in derivatives. This ensures that, when expanded around flat space, the theory is free of unphysical ghosts and instabilities.

A compelling approach to quantum gravity completion at high energies and its corresponding admissible low-energy effective theories is provided by the Swampland program. 
The Swampland conjectures, and in particular the Distance Conjecture, combined with the observed smallness of the cosmological constant,  $\Lambda \sim 10^{-120} M_{\text{Pl}}^4$,  imply the existence of one extra ``dark'' dimension, decompactified at the micron scale,  $R \sim 1\,\mu{\rm m}$~\cite{Montero:2022prj}. The Distance Conjecture requires that any large displacement in moduli space be accompanied by an infinite tower of states becoming light. Identifying this tower with the Kaluza–Klein modes of the dark dimension then correlates  the size of the extra dimension to the cosmological constant, fixing the radius  $R$ at the micron scale.

  In this context, with one dark dimension at the micron scale, the fundamental five-dimensional Planck scale is roughly $M_{*}\sim 10^{10}$GeV. Remarkably, combining this higher-dimensional framework with the EGB gravity can lead to significant modifications of black-hole physics. In four-dimensional General Relativity, Hawking evaporation causes the black-hole temperature to increase as the mass decreases, eventually reaching a regime in which the semiclassical description is no longer reliable. This signals the need for a quantum-gravitational description of the final stages of evaporation and is closely related to the black-hole information problem. In a higher-dimensional bulk governed by EGB gravity, higher-curvature terms modify the spacetime geometry and black-hole thermodynamics already at the classical level. In particular, the thermodynamics of the Boulware-Deser solutions~\cite{Boulware:1985wk}  receives corrections to the Bekenstein-Hawking area law through the Wald entropy, which can substantially modify the final stages of evaporation. In the the Dark Dimension scenario considered here, the Hawking temperature reaches a finite maximum rather than diverging as the black-hole mass decreases. Beyond this point, the thermodynamic behavior changes, and the evaporation can approach , a cold remnant with finite, non-zero mass.
    
    \paragraph{}On the other hand, in the case of two dark dimensions, the fundamental higher-dimensional Planck scale $M_{}$ is lowered to a bulk scale of roughly $M_{}\sim 10~\text{TeV}$. In this case, the heat capacity of the black hole remains negative, while the Hawking temperature diverges, $T_H\propto 1/r_h\to\infty$, as $r_h\to0$, precluding the formation of a classical cold remnant. Nevertheless, Gauss-Bonnet corrections modify the Stefan-Boltzmann evaporation rate and can significantly alter the black-hole lifetime. We further show that the heat capacity remains negative for any number of extra dimensions, $D>5$. This property is particularly notable in the context of the Dark Dimension scenario, as it distinguishes its black-hole thermodynamics from that of other higher-dimensional scenarios.

	In this paper, we explore the cosmological and phenomenological implications of EGB gravity corrections within the Dark Dimension scenario. We calculate how the Gauss-Bonnet coupling constant $\alpha$, linked to the fundamental string length $\ell _{s}$, shifts the horizon dynamics and alters the thermal properties of microscopic black holes. We analyze in detail the phase structure associated with a single decompactified extra dimension, $D=5$, while we  derive the corresponding mass and overabundance constraints for the resulting stable black-hole remnants.

Next, we calculate the lifetimes of microscopic 6D black holes and consider possible experimental signatures at future colliders. We also compare their thermodynamic properties with those of the five-dimensional solutions and identify the important differences that emerge at radii well below the Gauss–Bonnet scale. Finally, we discuss dynamical instability bounds and set out the conditions and physical requirements under which the remnant can be regarded as a viable, stable, cold dark matter candidate.

The layout of the paper is as follows: In Section~\ref{sec:EGB_D_Gravity}, we review the general $D$-dimensional formulation of EGB gravity, deriving the Boulware-Deser metric, Wald entropy corrections, and general heat capacity criteria across arbitrary dimensions. In Section \ref{sec:EGB_Gravity_Corrections_1DD}, we analyze the 5D Dark Dimension scenario ($D=5$), detailing the maximum temperature bound, the positive heat capacity branch, and the formation of stable cold remnants. In Section~\ref{sec:EGB_Gravity_Corrections_2DD}, we extend the analysis to two Dark Dimensions ($D=6$), deriving the corrected Stefan-Boltzmann evaporation power and the modified lifetime for macroscopic 6D black holes. In Section~\ref{sec:EGB_dynamical_stability}, we examine the linear gravitational-instability bounds of Beroiz, Dotti, and Gleiser (BDG) in five and six dimensions, contrasting the scalar- and tensor-instability mass thresholds with our thermodynamic branches and discussing the interplay between perturbation growth timescales, evaporation rates, and the limits of the stationary effective description. In Section~\ref{sec:validity_harmonics}, we analyze the ultraviolet (UV) cutoff of the truncated Einstein--Gauss--Bonnet effective field theory set by the fundamental string scale $l_s$, showing how sub-stringy wavelength constraints suppress high-harmonic scalar instabilities in $D=5$ and delineate the effective benchmark regime for microscopic black holes in $D=6$. In Section~\ref{sec:Experimental_Signatures}, we evaluate the cosmological viability of 5D EGB relics such as cold dark matter, establishing the required initial collapse fraction $\beta$ and the peak of the inflationary power spectrum $\mathcal{P}_\zeta$ to achieve $f_{\text{remnant}} = 1.0$ and discuss overabundance constraints. We explore the observational imprints of 5D relics, including scalar-induced gravitational wave backgrounds, suppression of evaporation bursts, and ultra-high-frequency gravitational waves from binary mergers. Also, we investigate microscopic 6D black holes: we establish their $\tau \propto M_0^3$ scaling law, compare $5\text{D}$ versus $6\text{D}$ thermodynamics, and discuss collider phenomenology. Finally, in Section \ref{sec:Conclusions}, we present our conclusions.

\section{Einstein-Gauss-Bonnet Gravity in D-dimensions}
\label{sec:EGB_D_Gravity}
	
	To set the stage, we start with the derivation of the basic black hole
	quantities in $D$-dimensional EGB gravity. In
	$D\ge 5$ dimensions, higher-derivative curvature terms such as the
	Gauss-Bonnet invariant modify black hole geometry and thermodynamics
	directly at the classical level.
	
	The $D$-dimensional EGB action is given by~\cite{Wald:1993nt}
	\begin{align}
		S_{\text{EGB}}
		=
		\frac{1}{16\pi G_D} \int d^D x \sqrt{-g}
		\left\{
		R + \alpha \left( R^2 - 4 R_{\mu\nu} R^{\mu\nu}
		+ R_{\mu\nu\rho\sigma} R^{\mu\nu\rho\sigma} \right)
		\right\},
	\end{align}
	where $\alpha$ is the Gauss-Bonnet coupling constant with dimensions of
	length squared, and $G_D$ is the $D$-dimensional Newton constant, defined in terms of the fundamental Planck scale $M_*$ as follows
	\begin{equation}
		G_D=\frac{1}{8\pi M_*^{D-2}}~.
		\label{GDdef}
	\end{equation}
We use this convention consistently throughout.
	
	The line element describing a higher-dimensional, neutral, static,
	spherically symmetric black hole in Gauss-Bonnet gravity is
	\begin{align}
		ds^{2}=h(r)\,dt^{2}-\frac{dr^{2}}{h(r)}-r^{2}d\Omega_{D-2}^{2},
	\end{align}
	where $d\Omega_{D-2}^{2}$ is the metric on a unit $(D-2)$-dimensional sphere,
	and the metric function $h(r)$ is given by~\cite{Boulware:1985wk}
	\begin{equation}
		h(r)=k+\frac{r^{2}}{2\alpha(D-3)(D-4)}
		\left(
		1\pm\sqrt{1+\frac{64\pi\alpha(D-3)(D-4)G_D M_{\rm BH}}
			{(D-2)\Omega_{D-2}r^{D-1}}}
		\right).
        \label{BoulwareMetric}
	\end{equation}
	Throughout we assume $D\ge 5$, so that the Gauss-Bonnet term is dynamically
	relevant; in $D=4$ it reduces to a total derivative and does not modify the
	field equations.
	
	Next, we consider the case of an asymptotically flat
	Schwarzschild-Gauss-Bonnet black hole, i.e.\ $k=1$, and take the $(-)$
	branch, while we comment on the $(+)$ Gauss-Bonnet branch later on. For
	later convenience, we define
	\begin{equation}
    		\label{definitions}
\begin{split}
	\mu& = \frac{16\pi G_D M_{\rm BH}}{(D-2)\Omega_{D-2}},\qquad\\
		\beta &= (D-3)(D-4)\alpha,\qquad\\
		W(r)&=\sqrt{1+\frac{4\beta\mu}{r^{D-1}}}.
        \end{split} 
	\end{equation}
	Then, from the horizon equation $h(r_+)=0$ we solve to find the condition
	\begin{equation}
		r_+^{D-3} + \beta\, r_+^{D-5} = \mu,
		\label{def:horequ}
	\end{equation}
	where $\beta$, defined in~(\ref{definitions}), is the rescaled Gauss-Bonnet coupling constant, with $\alpha\ge 0$.
	
	Furthermore, using~(\ref{definitions}) we can write the metric function $h(r)$
	in the simplified form
	\begin{equation}
		h(r)=1+\frac{r^{2}}{2\beta}\left(1-W(r)\right).
		\label{def:h_of_W}
	\end{equation}
	
	In $D$-dimensional EGB gravity, the radius of the horizon $r_h$ is defined as the
	largest real root of the metric function $h(r_h=r_+)=0$. Then, using
	(\ref{def:h_of_W}) we arrive at the following useful identity:
	\begin{equation}
		W(r_+) =1+\frac{2\beta}{r_+^2}.
		\label{def:Wfnction}
	\end{equation}
	
	For a static, spherically symmetric black hole, the Hawking temperature is
	$T_H=\kappa/(2\pi)$, where $\kappa$ is the surface gravity, which is expressed
	in terms of the derivative of the metric function at $r=r_+$. In particular,
	$\kappa = h'(r_+)/2$, hence
	\begin{equation}
		T_H=\frac{h'(r_+)}{4\pi}.
		\label{def:TH}
	\end{equation}
	
	We now compute $h'(r_+)$ explicitly. Differentiating $W(r)$ gives
	\begin{equation}
		W'(r)=-\frac{2\beta\mu (D-1)}{r^{D}W(r)}.
	\end{equation}
	Next,  differentiating (\ref{def:h_of_W}), we obtain
	\begin{equation}
		h'(r)=\frac{r}{\beta}\left(1-W(r)\right)
		-\frac{r^{2}}{2\beta}W'(r)
		=\frac{r}{\beta}\left(1-W(r)\right)
		+\frac{\mu(D-1)}{r^{D-2}W(r)}.
	\end{equation}
	Evaluating at $r=r_+$ and using the identity (\ref{def:Wfnction}) to eliminate
	$W(r_+)$, together with the horizon equation (\ref{def:horequ}) written as
	$\mu=r_+^{D-5}(r_+^2+\beta)$, we obtain
	\begin{equation}
		h'(r_+)
		=-\frac{2}{r_+}
		+\frac{(D-1)\left(r_+^2+\beta\right)}
		{r_+\left(r_+^2+2\beta\right)}
		=\frac{(D-3)r_+^2+(D-5)\beta}
		{r_+\left(r_+^2+2\beta\right)}~\cdot 
	\end{equation}
	Substituting the above into the temperature formula (\ref{def:TH}), yields,
	\begin{equation}
		T_H = \frac{(D-3) r_+^2 + (D-5) \beta}
		{4\pi r_+ \left(r_+^2 + 2\beta\right)}~\cdot 
		\label{HawkingTemp}
	\end{equation}
	
	It is reassuring to see that in the limit $\beta\to 0$ (i.e.\ $\alpha\to 0$)
	the standard Schwarzschild-Tangherlini\cite{Tangherlini:1963bw} solution of D dimensions is obtained.
	Two basic checks are shown here, while a detailed checklist is summarized in Table~\ref{tab:EGB_thermodynamics_general_vs_5D}. First, in the Schwarzschild limit $\beta\to 0$, the horizon
	equation gives $\mu\to r_+^{D-3}$, and
	\begin{equation}
		T_H \to \frac{D-3}{4\pi r_+},
	\end{equation}
	which is the standard $D$-dimensional Schwarzschild temperature. Second, in
	$D=5$ the term proportional to $\beta$ in the numerator of~(\ref{HawkingTemp})
	vanishes, and the temperature reduces to
	\begin{equation}
		T_H = \frac{r_+}{2\pi\left(r_+^2+2\beta\right)},
	\end{equation}
	which is the known five-dimensional EGB result.
	
	In Table~\ref{tab:EGB_thermodynamics_general_vs_5D} we summarize the successfully reproduced properties
	of the $D$-dimensional Schwarzschild-Tangherlini black hole in the limit
	$\alpha\to 0$ where the Gauss-Bonnet term is neglected.

	\begin{table*}[t]
\centering
\resizebox{\textwidth}{!}{%
\setlength{\tabcolsep}{8pt}
\renewcommand{\arraystretch}{1.6}
\begin{tabular}{lcc}
\hline\hline
\textbf{BH Physical Property} & \textbf{General $D$ Dimensions ($D \ge 5$)} & \textbf{Five Dimensions ($D=5$)} \\
\hline
Horizon Area $A_H$ & $\Omega_{D-2}\, r_h^{D-2}$ & $2\pi^2 r_h^3$ \\

Metric Horizon Condition $f(r_h)=0$ & $\mu = r_h^{D-3} \left(1 + \frac{{\beta}}{r_h^2}\right)$ & $\mu = r_h^2 + 2\alpha$ \\

ADM Mass $M(r_h)$ & $\frac{(D-2)\Omega_{D-2}}{16\pi G_D} \left(r_h^{D-3} + {\beta} r_h^{D-5}\right)$ & $\frac{3\pi}{8 G_5} r_h^2 + \frac{3\pi\alpha}{4 G_5}$ \\

Remnant Mass $M_{\text{rem}} \; (r_h \to 0)$ & $\frac{(D-2)\Omega_{D-2}}{16\pi G_D} \lim\limits_{r_h \to 0} \left[r_h^{D-3} + {\beta} r_h^{D-5}\right]$ & $\frac{3\pi\alpha}{4 G_5} > 0$ \\

Hawking Temperature $T_H(r_h)$ & $\frac{(D-3)r_h^2 + (D-5){\beta}}{4\pi r_h (r_h^2 + 2{\beta})}$ & $\frac{r_h}{2\pi (r_h^2 + 4\alpha)}$ \\

Asymptotic $T_H \; (r_h \to 0)$ & $\lim\limits_{r_h \to 0} T_H(r_h)$ & $0$ \\

Wald Entropy $S(r_h)$ & $\frac{A_H}{4 G_D} \left[1 + \frac{2(D-2)(D-3)\alpha}{r_h^2}\right]$ & $\frac{\pi^2 r_h^3}{2 G_5} + \frac{6\pi^2\alpha}{G_5} r_h$ \\

Heat Capacity $C(r_h)$ & $\frac{(D-2)\Omega_{D-2} r_h^{D-4}(r_h^2 + 2{\beta})^2 [(D-3)r_h^2 + (D-5){\beta}]}{4 G_D [-(D-3)r_h^4 + (9-D){\beta} r_h^2 - 2(D-5)\beta^2]}$ & $\frac{3\pi^2 r_h (r_h^2 + 4\alpha)^2}{2 G_5 (4\alpha - r_h^2)}$ \\

Thermodynamic Stability & Stable ($C > 0$) for $r_h < r_c$ only if $D=5$ & $C > 0$ for $r_h < 2\sqrt{\alpha}$ \\
\hline\hline
\end{tabular}%
}
\vspace{0.25cm}
\caption{Geometric and thermodynamic properties of spherically symmetric black holes in Einstein-Gauss-Bonnet gravity in arbitrary dimensions $D \ge 5$ compared to the specific case $D=5$ ($\beta = 2\alpha$). Here $\Omega_{D-2} = 2\pi^{(D-1)/2}/\Gamma[(D-1)/2]$ denotes the area of the unit $(D-2)$-sphere, and $\beta\equiv (D-3)(D-4)\alpha$.}
\label{tab:EGB_thermodynamics_general_vs_5D}
\end{table*}

	\subsection{Wald's entropy correction}

    Due to the presence of the Gauss–Bonnet term, higher-order curvature terms emerge, and consequently the standard Bekenstein–Hawking area law, 
$S = A/4G_D$, no longer holds. 
Wald's formula states that the entropy is given by the Noether charge of the horizon Killing vector. If the action depends on the Riemann tensor, the entropy also receives a contribution from the variation of the Lagrangian with respect to curvature. Applying Wald's Noether-charge formula for black hole entropy then yields~\cite{Jacobson:1993xs}
	\begin{align}
		S= \frac{A_H}{4 G_D} \left[ 1 + \frac{2(D-2)(D-3)\alpha}{r_h^2} \right],
		\label{WaldEnt}
	\end{align}
	where $A = \Omega_{D-2} r_h^{D-2}$ is the $(D-2)$-dimensional horizon area,
	$r_h$ is the horizon radius, and $k = +1, 0, -1$ denotes the spherical, flat, or
	hyperbolic horizon topology, respectively (see the Appendix for details). Thus, the GB invariant contributes an 
    extra term proportional to the integral of the intrinsic curvature of the horizon, 
    which for a spherical horizon reduces to the $\alpha/r_h^2$ correction in Eq.~(\ref{WaldEnt}).
  
    As with the temperature, in the  formula (\ref{WaldEnt}) it is assumed that $D\ge 5$, while  
    in $D=4$ the Gauss-Bonnet contribution to the
	entropy is absent. For standard spherical topology ($k = 1$), expanding the
	terms gives
	\begin{align}
		S_{EGB} = \frac{\Omega_{D-2}}{4G_D}r_h^{D-2} + \frac{\alpha(D-2)(D-3)\Omega_{D-2}}{2G_D} r_h^{D-4}
        \equiv \frac{\Omega_{D-2}}{4G_D} r_h^{D-2} \left(1+ \frac{D-2}{D-4}\frac{2\beta}{r_h^{2}}\right).
	\end{align}
As already pointed out, 
the GB term is a total derivative in four dimensions and therefore does not contribute 
to the equations of motion. Consequently,  the entropy correction vanishes in $D=4$, 
and the Bekenstein-Hawking area law is recovered.  Hence,  the  GB term is 
dynamically relevant only for $D\ge 5$ and the formula~(\ref{WaldEnt}) 
is therefore valid only for $D\ge 5$. Furthermore, we observe that for $\alpha > 0$ 
and $k=1$, the GB correction is positive and the entropy is larger than the area law,
whilst, as $\alpha\to 0$, the correction vanishes as expected.

\subsection{The Heat Capacity of the Black Hole}

A key quantity in black hole thermodynamics is the heat capacity $C$ of the
black hole. It determines whether the black hole is thermodynamically stable
and controls how the temperature responds to changes in mass.

In the remainder of this section we generalise the mass $M_{\rm BH}$ and the temperature $T_H$ of the black
hole to $D$ dimensions, and derive the heat capacity in $D$ dimensions.
These formulas will be useful in the following sections, where various
thermodynamic properties and the lifetime of EGB black holes are examined.

The $D$-dimensional expressions for the mass and temperature are
\begin{align}
    M_{\rm BH}
    &=
    \frac{(D-2)\Omega_{D-2}}{16\pi G_D}
    \left(r^{D-3}+\beta r^{D-5}\right),\\
    T_H
    &=
    \frac{(D-3)r^2+(D-5)\beta}
    {4\pi r\left(r^2+2\beta\right)}.
    \label{HeatCapD}
\end{align}
From these, we derive the heat capacity
$C=\dfrac{dM}{dT}=\dfrac{dM/dr}{dT/dr}$. The result is
\begin{equation}
    C
    =
    -\frac{(D-2)\Omega_{D-2}r^{D-4}}{4G_D}
    \frac{\left(r^2+2\beta\right)^2
    \left[(D-3)r^2+(D-5)\beta\right]}
    {(D-3)r^4+(D-9)\beta r^2+2(D-5)\beta^2}.
    \label{HeatCapacityD}
\end{equation}

Some useful remarks are in order.

 The denominator of the heat capacity,
     \begin{equation}
  {\cal D}  =    (D-3)r^4+(D-9)\beta r^2+2(D-5)\beta^2~,
        \label{Dtransition}
    \end{equation}
 contains the term $(D-9)\beta r^2$, which can be negative for $D<9$. 
 Consequently, it may change the sign of the denominator, 
    thereby turning the heat capacity from negative to positive and giving rise
    to a stable branch. For $D\ge 9$ the denominator (\ref{Dtransition}) 
    is positive for all $r>0$, so $C<0$ always.
   A negative heat capacity  means that the black hole is  thermodynamically unstable. 
   As the BH  loses mass, its temperature increases, accelerating further mass loss. In contrast, a positive heat capacity ($C>0$)
    means that the black hole is stable and, as it loses mass, its temperature
    falls, thus evaporation slows down and eventually stalls. The transition
    between the two regimes occurs where the denominator of $C$ vanishes, i.e.\ where $dT/dr=0$,
    which is exactly where the temperature reaches   its maximum.

 As $\beta\to 0$ (i.e.\ $\alpha\to 0$), the heat capacity is reduced to
    \begin{equation}
        C \to -\frac{(D-2)\Omega_{D-2}r^{D-2}}{4G_D}
        = -\frac{(D-2)A}{4G_D},
    \end{equation}
    which is negative for all $r>0$, as expected for a Schwarzschild black
    hole in $D$ dimensions. The stable branch disappears in this limit.

 As we pointed out above,  the  flip of the  sign of $C$ occurs at the radius of the horizon 
 where the  denominator vanishes.
    For general $D$, the transition radius is the positive root of ${\cal D} =0$, 
    where  ${\cal D} $  is the  denominator  defined in~(\ref{Dtransition}). 
    This radius sets the boundary between the unstable and stable branches.
 The stable branch exists only in $D=5$, while for all integers $D\ge 6$,
    the discriminant of the quartic equation~(\ref{Dtransition}) in the
    variable $r^2$ is negative, so there is no real positive root. The
    denominator of $C$ is therefore positive for all $r>0$, and $C<0$
    everywhere. In $D=6$, for example, the denominator becomes
    $3r^4-3\beta r^2+2\beta^2$, whose discriminant is $-15\beta^2<0$.
    This is consistent with the observation that in $D\ge 6$ the
    temperature diverges as $r_h\to 0$, whereas only in $D=5$ does it
    vanish, allowing a stable remnant to form.

In the following sections, we will reduce these equations for the specific cases of 
five and six dimensions and present details of the above properties and features 
in the context of Dark Dimension scenario and its six-dimensional analogue.

	\section{EGB gravity correction in the Dark Dimension scenario}
    \label{sec:EGB_Gravity_Corrections_1DD}
	
	For Dark Dimension (DD) scenarios we restrict the $D$-dimensional metric
	to the case $D=5$, thus we take
	\begin{equation}
    \beta = 2\alpha
		\qquad
		\mu = \frac{8G_5M}{3\pi}.
	\end{equation}
	Substituting into (\ref{BoulwareMetric}), taking $ D=5, k=1$ and changing the notation $h(r)\to f(r)$, for the $(-1)$ branch we obtain
	\begin{equation}
		f(r) = 1 + \frac{r^2}{4\alpha}
		\left(
		1 - \sqrt{1 + \frac{64 G_5 M \alpha}{3\pi r^4}}
		\right),
		\label{BDmetric}
	\end{equation}
	which is the metric function $f(r)$ for the spherically symmetric five-dimensional Boulware-Deser solution.
	
	Setting $f(r_h) = 0$ yields the explicit relation between the Arnowitt–Deser–Misner (ADM) mass $M$
	and the horizon radius $r_h$:
	\begin{equation}
		M(r_h) = \frac{3\pi}{8 G_5} \left( r_h^2 + 2\alpha \right).
		\label{MassBH5D}
	\end{equation}
	
	We use surface gravity to find the relation for the Hawking temperature:
	\begin{equation}
		T(r_h) = \frac{f'(r_h)}{4\pi}
		= \frac{r_h}{2\pi \left( r_h^2 + 4\alpha \right)}.
		\label{T5D}
	\end{equation}
	From the relation~(\ref{WaldEnt}) we find that the black hole entropy in
	$D=5$ is
	\begin{equation}
		S(r_h) = \frac{\pi^2}{2 G_5} \left( r_h^3 + 12 \alpha r_h \right),
		\label{S5D}
	\end{equation}
	where we used the horizon area $A_{D=5} = 2\pi^2 r_h^3$ and set $k=1$.
	
\begin{figure}[h]
		\centering
	\includegraphics[width=0.8\textwidth]{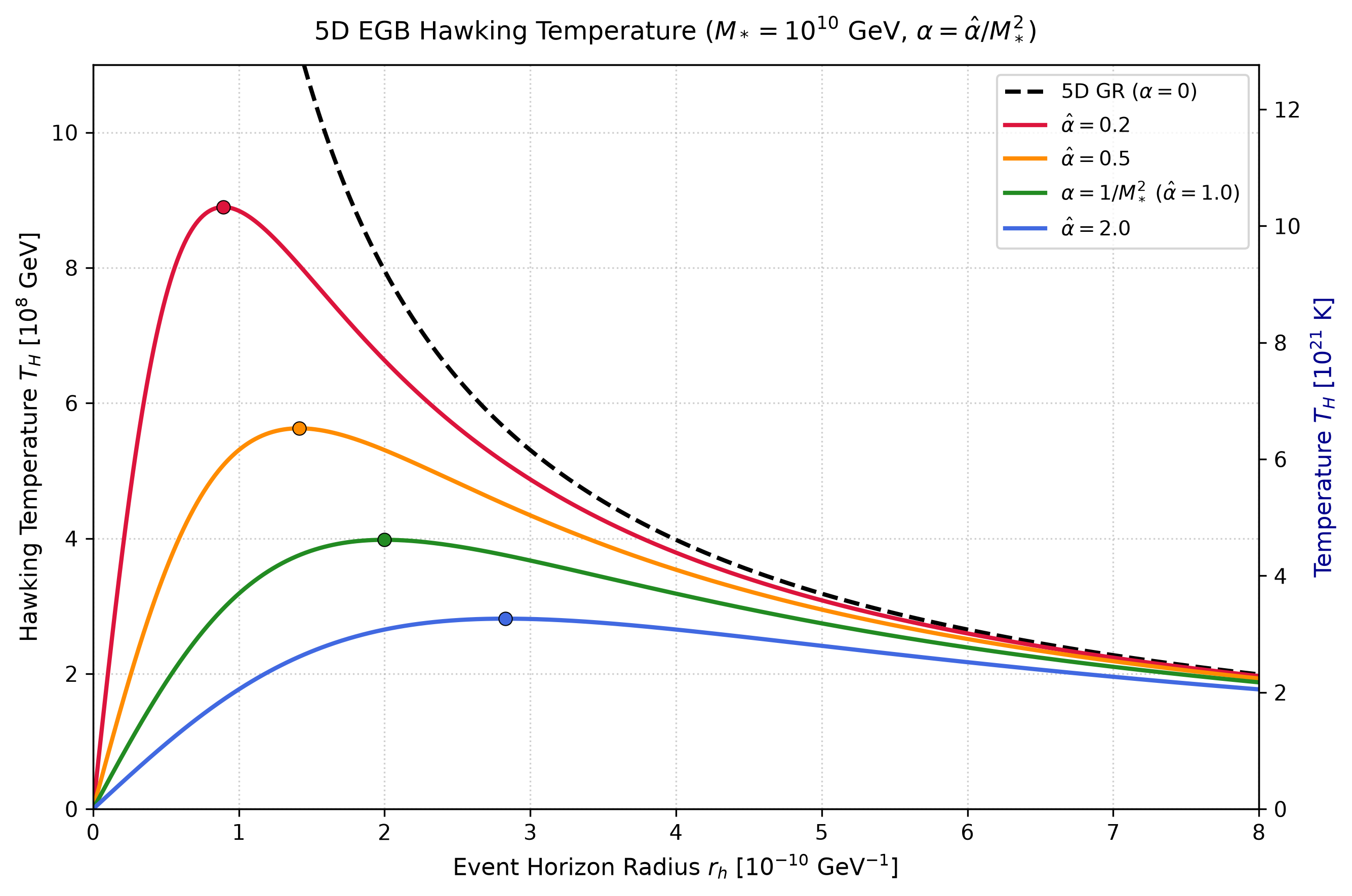}
		\caption{Plot of the temperature for various values of the GB coupling $\alpha$, Since $T_{\rm max}  \propto \alpha^{-1/2}$, as $\alpha$ decreases, $T_{\rm max} $ increases.}
		\label{5DTr}
	\end{figure}

	Differentiating the mass with respect to the temperature gives the heat
	capacity of the five-dimensional black hole~\cite{Cai:2001dz}:
	\begin{equation}
		C(r_h)
		=
		\frac{dM}{dT}
		=
		\frac{dM/dr_h}{dT/dr_h}
		=
		\frac{3\pi^2 r_h \left( r_h^2 + 4\alpha \right)^2}
		{2 G_5 \left( 4\alpha - r_h^2 \right)}.
		\label{C5D}
	\end{equation}
	We observe that, unlike the standard five-dimensional Schwarzschild black
	hole where $T \propto 1/r_h$ diverges as $r_h\to 0$
	(see Figure~\ref{5DTr}), the EGB temperature reaches a finite maximum
	at $r_h = 2\sqrt{\alpha}$:
	\begin{equation}
		T_{\max} = \frac{1}{8\pi\sqrt{\alpha}}.
		\label{Tpeak}
	\end{equation}

    The sign of heat capacity $C$ is controlled by the denominator in (\ref{C5D}). For $r_h < 2\sqrt{\alpha}$ the denominator is positive and the heat capacity  positive, $C > 0$. Thus the black hole stabilises thermally, cooling
	down as it loses its remaining mass.
	For $r_h > 2\sqrt{\alpha}$, the heat capacity  flips
	sign and becomes  negative  ($C < 0$). This  renders
	the black hole thermodynamically unstable, whilst  evaporation causes $T$ to rise.
	Finally, at the borderline between these two regions, $r_h = 2\sqrt{\alpha}$, the heat capacity diverges, marking a second-order phase transition.

	From Eq.~(\ref{MassBH5D}), as $r_h\to 0$ the Hawking temperature drops to zero,
	$T\to 0$. Instead of disappearing completely, evaporation 
    decreases and eventually stops, leaving
	behind a stable, cold remnant with a finite non-zero mass:
	\begin{equation}
		M_{\rm remnant}=\frac{3\pi\alpha}{4G_5}.
		\label{Mrem}
	\end{equation}
	Similarly, the radius of the horizon  as a function of the mass is
	\begin{equation}
		r_h(M) = \sqrt{\frac{8 G_5 M}{3\pi} - 2\alpha}.
		\label{rhofM}
	\end{equation}
	
	In low-energy effective field theories derived from string theory (such as
	heterotic or Type II superstring theory), $\alpha$ is explicitly set by the
	fundamental string length $\ell_s$:
	\begin{equation}
		\alpha = \frac{\alpha'}{8} = \frac{\ell_s^2}{8}.
		\label{alphastring}
	\end{equation}
	If scaled to the five-dimensional fundamental Planck length $\ell_5$, its
	magnitude is of order
	\begin{equation}
		\alpha \sim \ell_5^2 \approx 10^{-52}\ \text{m}^2.
	\end{equation}
	The value of $\alpha$ defines the boundary between classical general
	relativity and stringy quantum dynamics. For length scales
	$r \gg \sqrt{\alpha}$, classical general relativity dominates. For length
	scales $r \sim \sqrt{\alpha}$, the Gauss-Bonnet terms dominate and enforce
	the lower bound of the remnant mass  $M_{\rm remnant} = 3\pi\alpha/(4G_5)$.

	In the Dark Dimension scenario with an extra decompactified dimension at the
	micron scale, we have $G_5 = 1/(8\pi M_*^3)$ with $M_*=10^{10}$~GeV$\approx 1.78\cdot 10^{-14}$g. Using
	$\alpha\sim 1/M_*^2$, the mass of the remnant black hole is then
	\begin{equation}
		M_{\text{remnant}} = \frac{3\pi\alpha}{4G_5} = \frac{3\pi M_*^{-2}}{4(8\pi M_*^3)^{-1}} = 6\pi^2 M_* \approx 5.92 \times 10^{11}\text{ GeV} \approx 1.06 \times 10^{-12}\text{ g}
		\label{Mremnum}
	\end{equation}
	The 5D EGB solution admits a positive heat capacity branch whose formal zero-radius limit has a finite mass and vanishing Hawking temperature. 
    Whether this endpoint represents a physically thermodynamically stable remnant requires a treatment of the regime in which the semiclassical approximation breaks down.
     We emphasize that  thermodynamic stability  must be distinguished from the dynamical one. Positive heat capacity means the stationary solution is a local thermal equilibrium, however, it says nothing about whether the background is stable under gravitational perturbations; certainly  both must be ensured for a stable remnant. A detailed consideration of these issues  will be pursued in sections \ref{sec:EGB_dynamical_stability} and~\ref{sec:validity_harmonics}, where the conditions  for 
     stable remnants and possible caveats will be discussed.
   
   With these forewarnings in mind, the 5D EGB remnants may be said to bypass traditional four-dimensional constraints on PBH evaporation.

	\section{EGB gravity corrections in the two Dark Dimension scenario}
	\label{sec:EGB_Gravity_Corrections_2DD}

In this section, we will extend our analysis to the case of two extra compact dimensions of $\mu$m size~\cite{Anchordoqui:2025nmb,Leontaris:2026kvu}.
Interestingly, it was shown that the six-dimensional Planck scale can be as low as  10 TeV which is several orders 
of magnitude lower than that of the five-dimensional case. This fact makes two Dark Dimension scenario  
phenomenologically appealing since its predictions can be tested in future experiments. 

    We will  analyse various thermodynamical properties of this scenario and subsequently we will
    compare the results with those of the five-dimensional predictions discussed in the previous sections. 
    
	\subsection{The lifetime of macroscopic 6D PBHs}One of the fundamental questions we wish to answer  is the lifetime of a 6D 
	black hole. To find the lifetime of macroscopic PBHs, we first need the
	corrected Stefan--Boltzmann law in higher dimensions,
	$P(r_h)\propto A_{D-2}T_H^D$~\cite{Emparan:2000rs}. For the six-dimensional
	case of interest, we have
	\begin{align}
		P(r_h)
		&= \mu A_4 T^6
		= \mu \left( \frac{8\pi^2}{3} r_h^4 \right)
		\left[ \frac{3(r_h^2 + 2\alpha)}{4\pi r_h (r_h^2 + 12\alpha)} \right]^6
		\nonumber \\
		&= \frac{3^5 \, \mu}{2 \, (4\pi)^4 }
		\frac{(r_h^2 + 2\alpha)^6}{r_h^2(r_h^2 + 12\alpha)^6}.
		\label{Power6D}
	\end{align}
	
	We consider macroscopic black holes, \(r_h \gg 2\sqrt{\alpha}\), where
	\(\alpha=M_*^{-2}\), and evaluate the lifetime integral
	\begin{equation}
		\tau(M_0) = \int_0^{M_0} \frac{dM}{P(M)},
	\end{equation}
	changing variables from mass \(M\) to horizon radius \(r_h\).
	
	The BH mass in the six-dimensional Gauss-Bonnet case is obtained by setting
	\(D=6\) in the general mass formula,
	\begin{equation}
		M(r_h) = \frac{2\pi}{3 G_6} \left( r_h^3 + 6\alpha r_h \right).
		\label{Mass6D}
	\end{equation}
	Differentiating with respect to \(r_h\) gives
	\begin{equation}
		\frac{dM}{dr_h} = \frac{2\pi}{G_6} \left( r_h^2 + 2\alpha \right).
		\label{dMdr6D}
	\end{equation}
	Similarly, from the general temperature formula, the temperature of a
	macroscopic six-dimensional black hole is
	\begin{equation}
		T(r_h) = \frac{3(r_h^2 + 2\alpha)}{4\pi r_h (r_h^2 + 12\alpha)}.
		\label{T6D}
	\end{equation}
	
	The lifetime of a macroscopic black hole is then
	\begin{align}
		\tau(r_{h,0})
		&= \int_0^{r_{h,0}} \frac{dM/dr_h}{P(r_h)}\,dr_h
		\nonumber \\
		&= \left(\frac{4\pi}{3}\right)^5 \frac{1}{G_6 \mu}
		\int_0^{r_{h,0}}
		\frac{(r_h^2 + 12\alpha)^6}{(r_h^2 + 2\alpha)^5}\,r_h^2\,dr_h.
		\label{tauintegral}
	\end{align}
	In the regime \(r_h \gg \sqrt{\alpha}\) we can use the first-order Taylor expansion \((1+x)^n\approx 1+nx\) and write the integrand as
	\begin{equation}
		\frac{dM/dr_h}{P(r_h)}
		\approx \left(\frac{4\pi}{3}\right)^5 \frac{1}{G_6 \mu}
		\left( r_h^4 + 62\alpha r_h^2 \right).
		\label{integrandapprox}
	\end{equation}
	Consequently, the lifetime becomes~\footnote{We have checked that the exact
		closed form of the integral yields the same result upon expansion.}
	\begin{align}
		\tau(r_{h,0})
		&\approx \left(\frac{4\pi}{3}\right)^5 \frac{1}{G_6 \mu}
		\int_0^{r_{h,0}} \left( r_h^4 + 62\alpha r_h^2 \right) dr_h
		\nonumber \\
		&\approx \left(\frac{4\pi}{3}\right)^5 \frac{1}{G_6 \mu}
		\frac{r_{h,0}^5}{5}
		\left( 1 + \frac{310}{3} \frac{\alpha}{r_{h,0}^2} \right).
		\label{tauapprox}
	\end{align}
	Defining the general-relativistic lifetime
	\begin{equation}
		\tau_{\text{GR}}(r_{h,0})
		= \left(\frac{4\pi}{3}\right)^5 \frac{1}{G_6 \mu}
		\frac{r_{h,0}^5}{5},
		\label{tauGR}
	\end{equation}
	we finally obtain the folowing 
	\begin{equation}
		\tau(r_{h,0})
		\approx \tau_{\text{GR}}(r_{h,0})
		\left( 1 + \frac{310}{3} \frac{\alpha}{r_{h,0}^2} \right).
		\label{taufinal}
	\end{equation}
	To compare with standard GR at the same initial mass \(M_0\), we express
	\(r_{h,0}\) in terms of the classical six-dimensional Schwarzschild radius
	\(r_{\text{GR}} = \left( 3 G_6 M_0 / (2\pi) \right)^{1/3}\):
	\begin{equation}
		\tau(M_0)
		\approx \tau_{\text{GR}}(M_0)
		\left[ 1 + \frac{280}{3} \frac{\alpha}{r_{\text{GR}}^2} \right]
		= \tau_{\text{GR}}(M_0)
		\left[ 1 + \frac{280}{3} \alpha
		\left( \frac{2\pi}{3 G_6 M_0} \right)^{2/3} \right].
		\label{tauM0}
	\end{equation}

\subsection{The lifetime of microscopic 6D PBHs}

We devote our last discussion to the particular case of microscopic black
holes, i.e.\ when the horizon radius is much smaller than the Gauss-Bonnet
scale, $r_h\ll 2\sqrt{\alpha}$. Although their properties follow directly from
the general analysis of the previous section, we treat them separately because
the formulas simplify considerably and because of the different physics
prospects for future experiments.

We start with the 6D mass formula~(\ref{Mass6D}). For small radii,
$r_h\ll\sqrt{\alpha}$, the cubic term is negligible compared with the linear
term, $r_h^3\ll 6\alpha r_h$. In this limit the mass is well approximated by
\begin{equation}
    M(r_h) \approx \frac{2\pi}{3 G_6} (6\alpha r_h)
    = \frac{4\pi \alpha}{G_6} r_h.
    \label{MiniMass6D}
\end{equation}
Neglecting $r_h$ compared with the terms involving $\alpha$ in
Eq.~(\ref{T6D}), the temperature for small $r_h$ becomes
\begin{equation}
    T(r_h) \approx \frac{3(2\alpha)}{4\pi r_h (12\alpha)}
    = \frac{1}{8\pi r_h}.
\end{equation}
Using Eq.~(\ref{MiniMass6D}) to express $r_h$ in terms of the mass, we obtain
\begin{equation}
    T(M) \approx \frac{1}{8\pi \left(G_6 M/(4\pi \alpha)\right)}
    = \frac{\alpha}{2 G_6 M}.
    \label{TofM6Dmini}
\end{equation}

The radiated power is
\begin{align}
    P(M)
    &= \mu \left( \frac{G_6^4 M^4}{96 \pi^2 \alpha^4} \right)
    \left( \frac{\alpha}{2 G_6 M} \right)^6
    = \frac{\mu \, \alpha^2}{6144 \pi^2 G_6^2 M^2}.
    \label{P6Dmini}
\end{align}
Substituting $P(M)$ into the lifetime integral,
\begin{align}
    \tau(M_0)
    &= \int_0^{M_0} \frac{dM}{\mu \, \alpha^2/(6144 \pi^2 G_6^2 M^2)}
    = \frac{6144 \pi^2 G_6^2}{\mu \, \alpha^2} \int_0^{M_0} M^2\,dM
    \nonumber \\ \\ \nonumber
    &= \frac{2048 \pi^2 G_6^2}{\mu \, \alpha^2} M_0^3.
    \label{tauEGBmini}
\end{align}
The ratio of the microscopic EGB lifetime to the GR lifetime is,
\begin{equation}
    \frac{\tau_{\text{EGB}}(M_0)}{\tau_{\text{GR}}(M_0)}
    =
    \frac{(2048 \pi^2 G_6^2/\mu\alpha^2) M_0^3}
    {\left(1024 \pi^{10/3}(3/2)^{5/3}/(1215\mu)\right)
    G_6^{2/3} M_0^{5/3}},
    \label{ratio6D}
\end{equation}
Thus, the lifetime of a microscopic 6D EGB black hole can be written as
\begin{equation}
   \tau_{\text{EGB}}/\tau_{\text{GR}} \approx 3.650\,(M_0/M_*)^{4/3}.
    \label{Life6Dmini}
\end{equation}

\begin{figure}[h]
    \centering
    \includegraphics[width=1.0\textwidth]{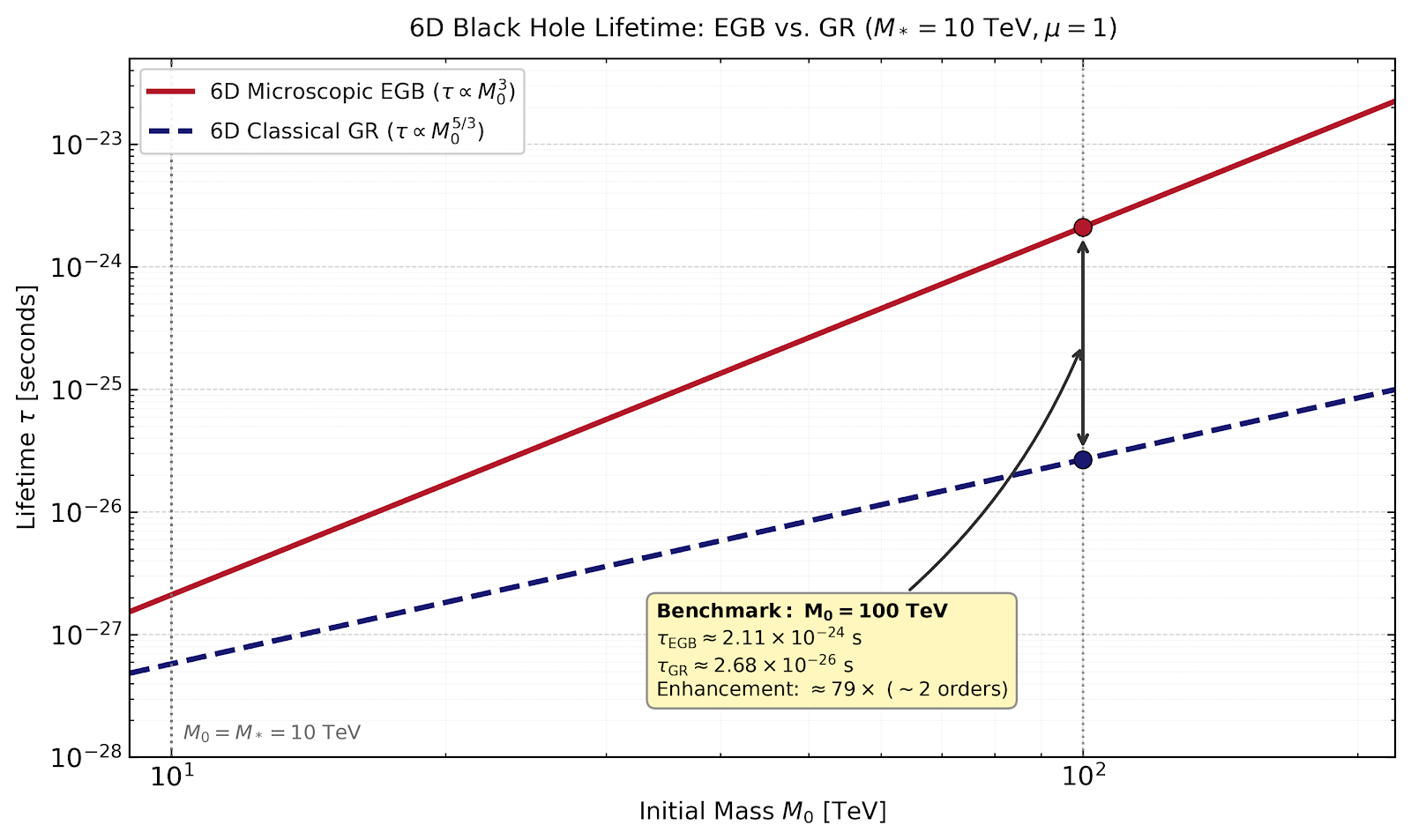}
    \caption{Comparison of black hole evaporation lifetimes $\tau$ (in
    seconds) as a function of initial mass $M_0$ (in TeV) in 6D spacetime,
    evaluated for a fundamental scale $M_*=10$~TeV and effective emission
    coefficient $\mu=1$. On a log--log scale, the classical 6D general
    relativity prediction follows $\tau_{\text{GR}}\propto M_0^{5/3}$,
    whereas the microscopic 6D EGB  model exhibits a
    steeper scaling $\tau_{\text{EGB}}\propto M_0^3$. Higher-curvature
    Gauss-Bonnet corrections suppress Hawking radiation at microscopic
    scales ($r_h\ll\sqrt{\alpha}$), extending the black hole lifetime by
    nearly two orders of magnitude at $M_0=100$~TeV compared with pure GR.}
    \label{PBHlifeA}
\end{figure}

Figure~\ref{PBHlifeA} illustrates the difference in lifetime between EGB and GR primordial black holes for various initial masses.  The steeper scaling $\tau_{\text{EGB}}\propto M_0^3$ may have implications
for high-energy collider physics. At microscopic scales
($r_h\ll\sqrt{\alpha}$), the higher-curvature Gauss-Bonnet terms act as a
thermodynamic buffer, significantly reducing the Hawking emission rate
relative to classical general relativity. For mini black holes
hypothetically produced at a future high-energy collider such as a
$100$~TeV Future Circular Collider, the EGB corrections could prolong their
active lifetime from about $10^{-26}$~s to about $10^{-24}$~s, an extension
of nearly two orders of magnitude. Such a longer lifetime could modify the
expected event topology, turning what would be an instantaneous, explosive
burst into a more prolonged decay that might be tracked inside the detector.

  \subsection{Comparison with the 5D black hole properties}

Let us compare the thermodynamic properties of black holes in five and six
dimensions. This is of particular interest because important differences
arise at small radii.

For large 6D black holes, $r_h\gg\sqrt{\alpha}$, the temperature is well
approximated by
\begin{equation}
    T \approx \frac{3}{4\pi r_h}.
\end{equation}
For $r_h\to\infty$, the temperature vanishes. This is the standard 6D
Schwarzschild behaviour.

For small black holes, $r_h\ll\sqrt{\alpha}$, the approximation yields
\begin{equation}
    T \approx \frac{1}{8\pi r_h}.
\end{equation}
As $r_h\to 0$, the temperature diverges, just as in the 4D Schwarzschild
case. The reason is that the mass term is dominated by the linear term in
$r_h$: for $D=6$ the term $\beta r_h^{D-5}$ becomes $\beta r_h$, whereas for
$D=5$ this term is absent because $D-5=0$.

Thus, in 6D  in the limiting case of small $r_h$ there is no stable remnant with
$T\to 0$. Instead, the temperature diverges, and the black hole continues to
evaporate. It is possible, however, that the semiclassical description breaks
down at such small radii and cannot give a reliable result.

More generally, the behaviour of the temperature as $r_h\to 0$ is controlled
by the coefficient $D-5$ of $\beta$ in the numerator of
\begin{equation}
    T_H = \frac{(D-3)r_h^2+(D-5)\beta}{4\pi r_h(r_h^2+2\beta)}.
\end{equation}
For $D=5$, this coefficient vanishes, so the numerator vanishes as $r_h^2$
and $T_H\to 0$. For $D>5$, the coefficient is positive, so the numerator
approaches a constant and $T_H\to\infty$. In the latter case, the
small-$r_h$ behaviour $T\propto 1/r_h$ is the same as in 4D Schwarzschild:
the Gauss-Bonnet term effectively reduces the near-horizon geometry to four
dimensions.

From this point of view, the case $D=5$ is special. The vanishing of the
numerator as $r_h\to 0$ occurs only for $D=5$, because the coefficient
$D-5$ of $\beta$ is zero in that case (see appendix~\ref{DD5}). This feature makes the $D=5$ case
particularly interesting, since it provides a simple mechanism by which the
temperature vanishes as $r_h\to 0$, so that a portion of the initial black
hole mass does not evaporate and could account for a fraction of the dark
matter in the universe.

We emphasize, however, that for $r_h\sim\sqrt{\alpha}\sim\ell_5$ the horizon
radius is of order the fundamental Planck or string length, and the
semiclassical approximation is not reliable. Any statement about the endpoint
of evaporation in $D\ge 5$ requires a full theory of quantum gravity.

	\begin{figure}[h]
		\centering
	\includegraphics[width=0.45\textwidth]{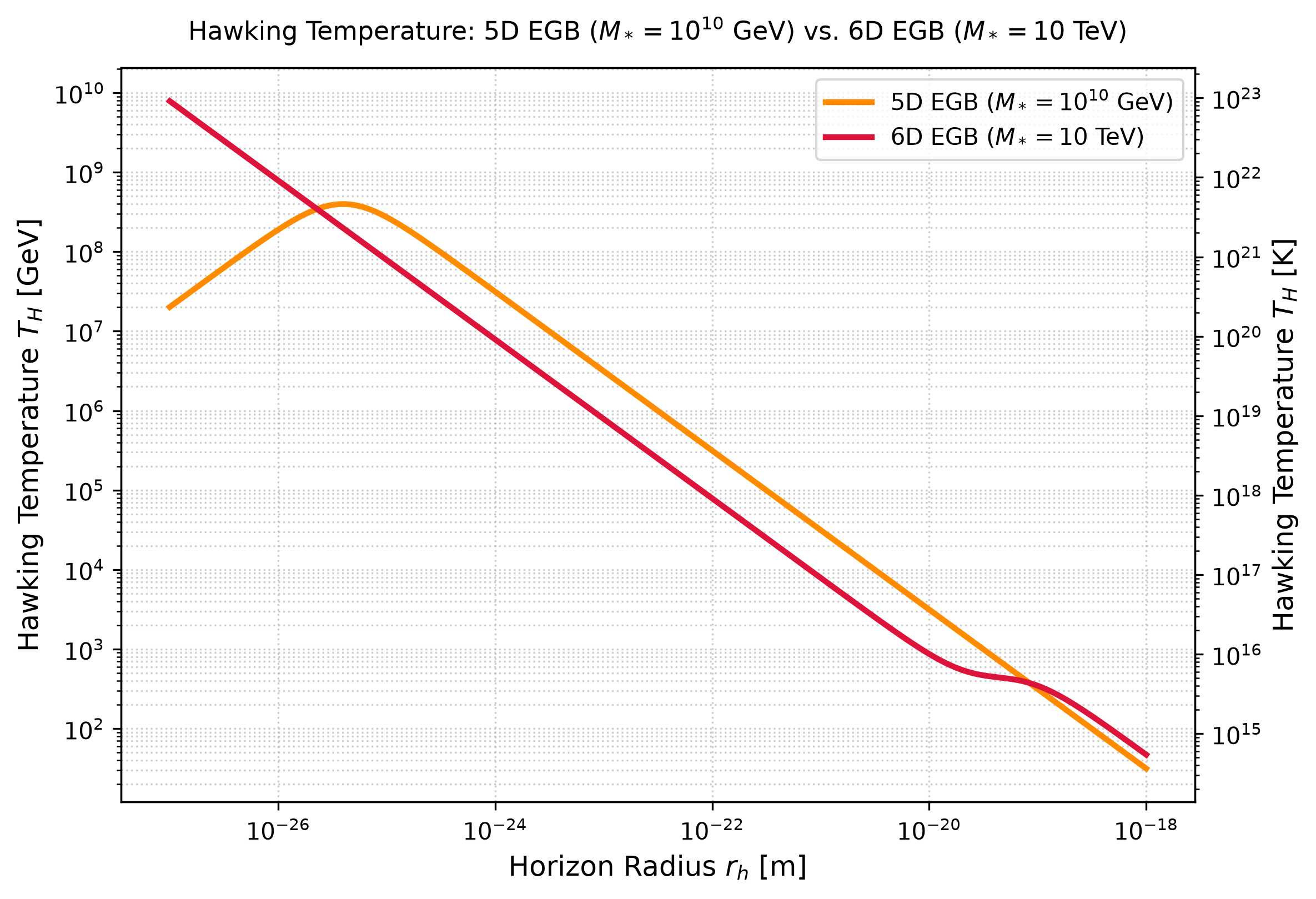}
    \includegraphics[width=0.45\textwidth]{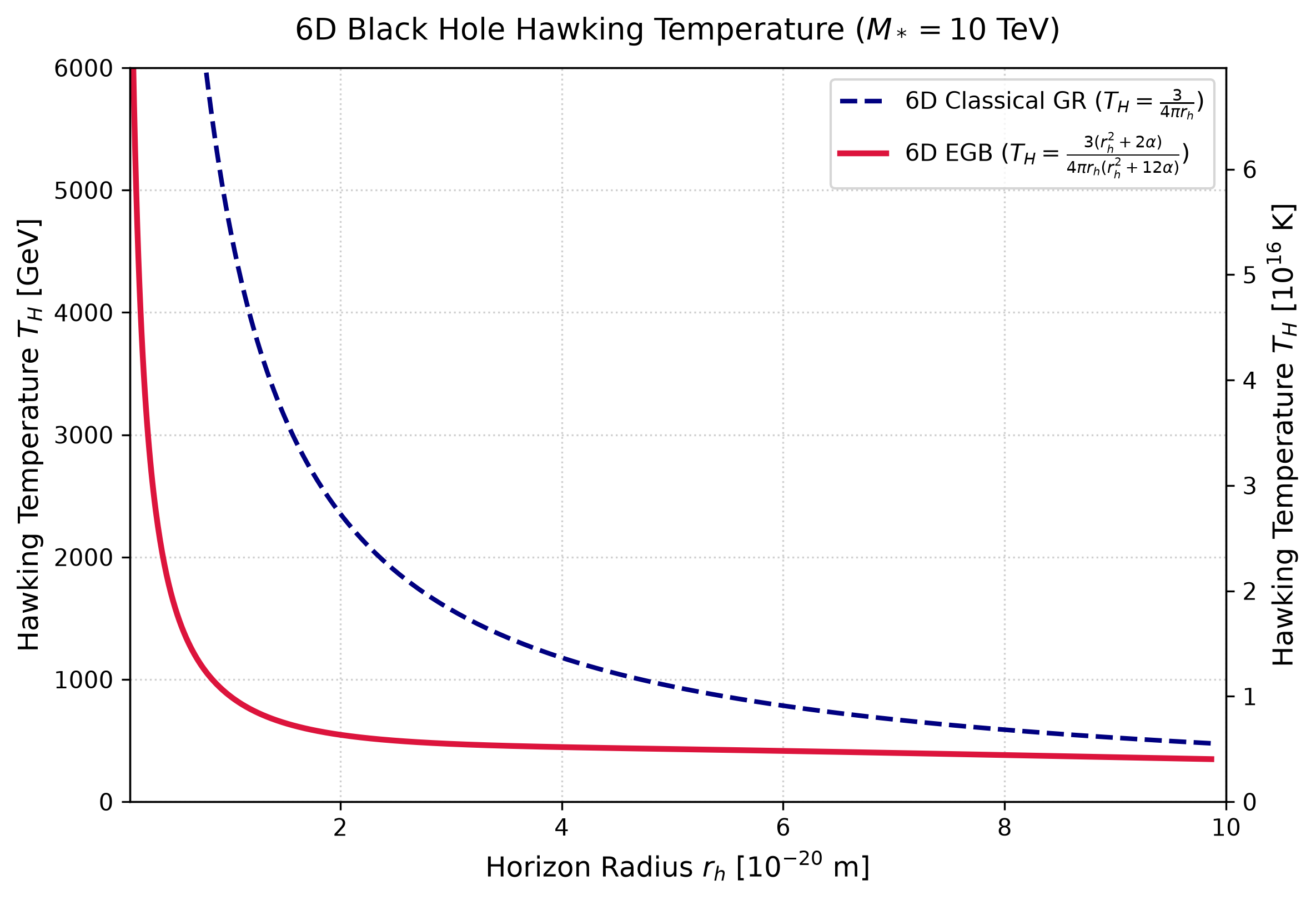}
		\caption{Left panel: Hawking temperature $T_H$ versus horizon radius $r_h$ in five and six dimensions in EGB gravity. Right panel: the same quantity compared between EGB and GR gravity.}
		\label{5D6Dcompare}
	\end{figure}

\section{Dynamical instability bounds in five- and six-dimensional EGB gravity}
\label{sec:EGB_dynamical_stability}
In the preceding sections, we demonstrated the existence of a stationary black-hole solution and showed that in five dimensions
its heat capacity can be positive within a specific mass range. Neither result, however, establishes dynamical stability.
To assess the interpretation of the thermodynamic branches considered above, we compare them with the linear gravitational-instability 
bounds obtained by BDG~\cite{Beroiz:2007gp}. Their analysis concerns neutral, static, spherically symmetric,
asymptotically flat black holes on the Einstein branch, with positive Gauss--Bonnet coupling. Its application to a compactified setting 
assumes that the horizon and the relevant perturbations probe scales sufficiently smaller than the compactification radii for the uncompactified solution to be a useful approximation. While these bounds are associated with  stationary backgrounds  they do not by themselves determine instability timescales. In the following, we examine the stability for the 5D and 6D cases separately.

In order to make contact with notation and conventions of ~\cite{Beroiz:2007gp}, we define $n=D-2$, and use the metric function
\begin{equation}
 f_{\rm BDG}(r)=1+
 \frac{r^2}{\alpha_{\rm BDG}(n-1)(n-2)}
 \left[1-\sqrt{1+
 \frac{4\alpha_{\rm BDG}(n-1)(n-2)\mu_{\rm BDG}}
 {n r^{n+1}}}\right].
 \label{eq:BDG_metric_general}
\end{equation}
In our convention, where the action contains $R+\alpha\mathcal{L}_{\rm GB}$,
\begin{equation}
 f(r)=1+\frac{r^2}{2\beta}
 \left[1-\sqrt{1+\frac{4\beta\mu}{r^{D-1}}}\right],
 \qquad \beta=(D-3)(D-4)\alpha,
 \label{eq:our_metric_stability}
\end{equation}
with
\begin{equation}
 \mu=\frac{16\pi G_D M}{(D-2)\Omega_{D-2}}.
\end{equation}
Matching the metric functions, gives the following relations
\begin{equation}
 \alpha_{\rm BDG}=2\alpha,
 \qquad \mu_{\rm BDG}=\frac{D-2}{2}\mu~.
 \label{eq:BDG_translation}
\end{equation}
The parameter $\mu$ has dimension $({\rm length})^{D-3}$,  therefore,  
its conversion to a dimensionless BDG mass differs between five and six dimensions.

\subsection{Instabilities in five dimensions}
\label{subsec:BDG_five_dimensions}
We will examine whether the positive-heat-capacity region determined in previous sections is dynamically unstable.
In $D=5$, the BDG defines
\begin{equation}
 m\equiv\frac{\mu_{\rm BDG}}{\alpha_{\rm BDG}}
 =\frac{3\mu}{4\alpha}
 =\frac{2G_5M}{\pi\alpha}.
\end{equation}
It is useful to introduce the formal limiting mass
\begin{equation}
 M_{\rm lim}\equiv\frac{3\pi\alpha}{4G_5},
 \qquad
 m=\frac32\frac{M}{M_{\rm lim}}.
 \label{eq:BDG_five_mass}
\end{equation}
The quantity $M_{\rm lim}$  is the mass obtained by taking  $r_h\to0^+$ along the stationary EGB family. 
Although it was previously denoted  $M_{\rm remnant}$  in Eq.~(\ref{Mrem}) of 
Sec.~\ref{sec:EGB_Gravity_Corrections_1DD}, as a matter of fact, an additional physical assumption is required before we can regard it as a remnant.

The horizon relation becomes
\begin{equation}
 r_h^2=\frac{8G_5M}{3\pi}-2\alpha
 =2\alpha\left(\frac{M}{M_{\rm lim}}-1\right).
 \label{eq:BDG_five_horizon}
\end{equation}
A finite, positive horizon radius thus requires $M>M_{\rm lim}$, or $m>3/2$. The equality $m=3/2$ is a singular zero-radius limit and should not be identified with a regular black hole having a finite horizon.

BDG prove the existence of a scalar-type gravitational instability at sufficiently high angular harmonic number in the interval
\begin{equation}
 \frac32<m<m_c^{(5)},
 \qquad m_c^{(5)}=\frac92+3\sqrt2\simeq8.743.
\end{equation}
In our variables, this interval is
\begin{equation}
 M_{\rm lim}<M<M_c^{(5)},
 \qquad
 \boxed{M_c^{(5)}=(3+2\sqrt2)M_{\rm lim}
 \simeq5.828M_{\rm lim}.}
 \label{eq:BDG_five_critical_mass}
\end{equation}
The corresponding critical radius is
\begin{equation}
 r_c^{(5)}=2\sqrt{(1+\sqrt2)\alpha}
 \simeq3.108\sqrt\alpha.
\end{equation}

This bound can be compared directly with the thermodynamic turning point. The heat capacity is positive for $0<r_h<2\sqrt\alpha$, which corresponds to
\begin{equation}
 1<\frac{M}{M_{\rm lim}}<3.
\end{equation}
Consequently, the entire positive-heat-capacity branch lies within the BDG scalar instability region. A decreasing-mass sequence of stationary solutions enters this region before reaching the temperature maximum at $r_h=2\sqrt\alpha$. Positive heat capacity therefore does not justify extrapolating a dynamically stable evaporation trajectory to $M_{\rm lim}$.

Notice, however, that the instability theorem applies to finite-radius backgrounds in
the open interval $M_{\rm lim}<M<M_c^{(5)}$, and not to the zero-radius endpoint
$M=M_{\rm lim}$ itself. Here $M_{\rm lim}=3\pi\alpha/(4G_5)$ is the formal limit of
the stationary Boulware-Deser family as $r_h\to0^+$ (the quantity denoted
$M_{\rm remnant}$ in Sec.~\ref{sec:EGB_Gravity_Corrections_1DD}), at which the horizon radius vanishes and the expansion
of the metric function degenerates; it is a singular \emph{limit of the family}, not a
statement that the endpoint is a curvature singularity, and the BDG linear perturbation
analysis, (which is performed on a finite-radius background), has nothing to say about
it. Consequently the theorem establishes neither that the black hole is completely
destroyed nor that it settles into a stable quantum or string remnant. It simply 
establishes that a quasi-static evaporation trajectory that passes through the
interval $r_c^{(5)}\gtrsim r_h>0$ is dynamically unstable, so the stationary
thermodynamic extrapolation of Sec.~\ref{sec:EGB_Gravity_Corrections_1DD} cannot be continued through that interval without
additional input. The fate of the endpoint is doubly inaccessible to the present
analysis: it lies outside the domain of the BDG theorem because it is a zero-radius
limit, and it lies outside the semiclassical approximation because
$r_h\sim\sqrt\alpha\sim\ell_s$. Determining
whether a remnant survives therefore requires a treatment of the regime in which the
truncated EGB effective action itself breaks down.

\subsection{Six dimensions: the instability threshold and collider benchmarks}
\label{subsec:BDG_six_dimensions}

In six dimensions ($n=4$), the translation yields $\alpha_{\text{BDG}} = 2\alpha$ and $\mu_{\text{BDG}} = 2\mu$. 
The dimensionless parameter of  BDG is defined as $m \equiv \mu_{\text{BDG}} / \alpha_{\text{BDG}}^{3/2}$,~\cite{Beroiz:2007gp} which translates to:
\begin{equation}
m = \frac{2\mu}{(2\alpha)^{3/2}} = \frac{1}{\sqrt{2}} \frac{\mu}{\alpha^{3/2}}.
\end{equation}
Using $\mu = \frac{3 G_6 M_{\text{BH}}}{2\pi}$ along with the fundamental 6D relations $G_6 = \frac{1}{8\pi M_*^4}$ and $\alpha = \frac{1}{M_*^2}$,
the parameter $m$ becomes:
\begin{equation}
m = \frac{1}{\sqrt{2} \left(M_*^{-2}\right)^{3/2}} \left( \frac{3 M_{\text{BH}}}{16\pi^2 M_*^4} \right) = \frac{3}{16\sqrt{2}\pi^2} \left( \frac{M_{\text{BH}}}{M_*} \right) \approx 0.0134 \left( \frac{M_{\text{BH}}}{M_*} \right).
\end{equation}
In $D=6$, BDG found that EGB black holes are dynamically unstable under tensor perturbations for $m < m_{c(6\text{D})} \approx 7.965$. Setting this threshold yields the critical stability mass:
\begin{equation}
0.0134 \left( \frac{M_{\text{BH}}^{\text{crit}}}{M_*} \right) \approx 7.965 \implies M_{\text{BH}}^{\text{crit}} \approx 593 \, M_*.
\end{equation}
For a fundamental scale of $M_* = 10\text{ TeV}$, the classical metric requires $M_{\text{BH}} \ge 5.93\times10^3\text{ TeV}$ to maintain a stable, spherical tensor configuration. Consequently, all microscopic black holes considered for future colliders ($M_{\text{BH}} \sim 100\text{--}500\text{ TeV}$, giving $M_{\text{BH}}/M_* \sim 10\text{--}50$) yield yield $m \approx 0.134\text{--}0.672 \ll 7.965$. They are deeply embedded in the tensor instability region.

Although the stationary metric exists below $M_c^{(6)}$, its Hawking temperature, geometric production area, and blackbody lifetime cannot automatically be interpreted as predictions for an approximately stationary object produced in a collision. Such an interpretation requires formation dynamics and a comparison between perturbation growth and evaporation. The instability threshold itself determines neither a production cross section nor a particle-decay spectrum.

    \section{On the validity of BDG limits and harmonics}
    \label{sec:validity_harmonics}

  The physical interpretation of the BDG instability requires care when EGB gravity is treated
  as a truncated string effective theory, since the instability may fall outside the regime
  in which the truncation is reliable. 
An angular harmonic of number $\ell$ has characteristic local momentum
$k_{\rm ang}\sim\ell/r_h$, corresponding  to an angular
wavelength 
\begin{equation}
\lambda_{\rm ang}\sim r_h/\ell~.\label{eq:wavelength}
\end{equation}
 The derivative expansion requires $k_{\rm ang}\ell_s\ll1$, or equivalently
$\lambda_{\rm ang}\gg\ell_s$, and therefore does not permit an
unrestricted extrapolation to arbitrarily high harmonics.
For each background, one must determine whether unstable modes occur
within this controlled range; high-harmonic instability alone does not
answer that question, and lower-harmonic instabilities must also be
examined. Independently, the background curvature must remain small
in string units. If these conditions fail, the truncated EGB theory
cannot reliably determine the evolution or its endpoint.

\subsection{Timescales, UV Cutoff, and Domain of Validity}

The dynamical instability bounds derived by BDG~\cite{Beroiz:2007gp} rely on the mathematical assumption of a strictly static vacuum background and on the existence of arbitrarily high angular harmonics ($\ell \gg 1$). However, treating EGB gravity as a low-energy effective field theory truncated at $\mathcal{O}(\alpha')$ imposes stringent physical limitations on the validity of this classical instability, particularly at the microscopic scales considered in the Dark Dimension scenario.

In the EFT framework, the Gauss--Bonnet coupling is fixed by the fundamental string length $l_s$ via Eq.~(\ref{alphastring}), $\alpha = l_s^2 / 8$, which implies $l_s = \sqrt{8\alpha}$. A linear perturbation characterized by an angular harmonic number $\ell$ corresponds to a transverse physical wavelength on the horizon of parametric size $\lambda_\ell \sim 2\pi r_h/\ell$. For the effective description to remain valid, this wavelength must not resolve sub-stringy distances. Modes with $\lambda_\ell \ll l_s$ lie strictly outside the physical Hilbert space of the truncated effective action and are regularized in the full theory by the infinite tower of higher-derivative string corrections ($\mathcal{O}(\alpha'^2)$) and by extended-object dynamics.

In $D=6$, the BDG bound indicates that static tensor perturbations ($\ell \ge 2$) become unstable for $M_{\text{BH}} \le 593\,M_*$. For the microscopic black holes expected at future colliders ($100\text{--}500\text{ TeV}$), the mass--radius relation is highly compressed. Using $\alpha = 1/M_*^2$ together with $l_s = \sqrt{8\alpha} = 2\sqrt{2}/M_*$, the horizon radius scales as
\begin{equation}
r_h \approx \frac{M_{\text{BH}}}{32\pi^2 M_*^2} = \frac{1}{32\pi^2}\left(\frac{M_{\text{BH}}}{M_*}\right)\frac{1}{M_*} = \frac{1}{64\sqrt{2}\pi^2}\left(\frac{M_{\text{BH}}}{M_*}\right) l_s \approx 0.00112\left(\frac{M_{\text{BH}}}{M_*}\right) l_s.
\end{equation}
For a given angular harmonic $\ell$ on the $S^4$ horizon, the physical wavelength is set by the transverse momentum. For symmetric transverse-traceless tensor perturbations\cite{Kodama:2003jz,Dotti:2005sq} on $S^4$, the Laplace--Beltrami eigenvalue is $\ell(\ell+3)-2$, so that
\begin{equation}
\lambda_\ell \approx \frac{2\pi r_h}{\sqrt{\ell(\ell+3)-2}}.
\end{equation}
The most dangerous mode, and the easiest to excite, is the lowest allowable tensor harmonic ($\ell = 2$). For this fundamental mode, the wavelength evaluated on the horizon is
\begin{equation}
\lambda_{\ell=2} = \frac{2\pi r_h}{\sqrt{2(2+3)-2}} = \frac{2\pi r_h}{\sqrt{8}} = \frac{\pi}{\sqrt{2}}\,r_h \approx 2.22\,r_h.
\end{equation}

Before we show a few representative numerical examples, a cautionary remark bears repeated here: 
for the assumed relation between the string length and Gauss–Bonnet coupling, the small-radius backgrounds and sufficiently short-wavelength perturbations lie outside the controlled derivative expansion. Consequently, the truncated EGB description establishes neither the ultraviolet persistence of the classical instability nor the stability of a surviving remnant.
Having this in mind, we proceed to an explicit evaluation for the critical collider mass benchmarks (assuming a fundamental scale $M_* = 10\text{ TeV}$).  We find:
\begin{itemize}
    \item At $M_{\text{BH}} = 100\text{ TeV}$ ($M/M_* = 10$): the horizon is $r_h \approx 0.0112\,l_s$, and $\lambda_{\ell=2} \approx 2.22 \times 0.0112\,l_s \approx 0.025\,l_s \ll l_s$.
    \item At $M_{\text{BH}} = 200\text{ TeV}$ ($M/M_* = 20$): $r_h \approx 0.0224\,l_s$, and $\lambda_{\ell=2} \approx 2.22 \times 0.0224\,l_s \approx 0.050\,l_s \ll l_s$.
    \item At $M_{\text{BH}} = 500\text{ TeV}$ ($M/M_* = 50$): $r_h \approx 0.0560\,l_s$, and $\lambda_{\ell=2} \approx 2.22 \times 0.0560\,l_s \approx 0.124\,l_s \ll l_s$.
\end{itemize}
These estimates show that across the entire accessible future collider energy domain ($M_{\text{BH}} \le 500\text{ TeV}$), the tensor instability would require the excitation of modes whose physical wavelengths are far smaller than the fundamental string length ($\lambda_{\ell=2} \ll l_s$). At such sub-stringy (trans-Planckian) resolutions the classical $\mathcal{O}(\alpha')$ EGB description breaks down. We stress, however, that the breakdown is not confined to the perturbation sector: at these masses the \emph{background} horizon itself satisfies $r_h \ll l_s$, so the stationary Boulware--Deser geometry is itself outside the domain of the truncated EGB effective action. The BDG bound is therefore not so much avoided as inapplicable, along with the entire classical EGB description of such objects. The few-body Quantum Black Hole (QBH) decay kinematics detailed in Section~7.2 should be read in this light: they are  effective phenomenological benchmarks near the boundary of the semiclassical regime, rather than controlled EFT predictions.

In $D=5$, the BDG scalar instability operates in the mass interval $M_{\text{rem}} < M \le 5.83\,M_{\text{rem}}$, corresponding to horizon radii $0 < r_h \le 3.108\sqrt{\alpha}$. The critical parameter $k_S = \sqrt{\ell(\ell+2)}$ controls the behaviour of the effective potential governing the perturbations. As shown in the BDG perturbation analysis, the negative-energy bound state responsible for the instability appears only at sufficiently high angular harmonic number ($\ell \gg 1$); the instability is an eikonal-type phenomenon, not a low-$\ell$ one.

Evaluating the physical wavelength of these high harmonics on the $S^3$ horizon yields
\begin{equation}
\lambda_\ell \approx \frac{2\pi r_h}{\ell}.
\end{equation}
To remain within the validity domain of the string effective field theory, the physical wavelength must exceed the string length ($\lambda_\ell \ge l_s = 2\sqrt{2\alpha}$). This imposes a dynamical ultraviolet cutoff on the maximum allowed harmonic number,
\begin{equation}
\ell_{\max}(r_h) \approx \frac{2\pi r_h}{l_s} = \frac{\pi}{\sqrt{2}}\,\frac{r_h}{\sqrt{\alpha}} \approx 2.22\,\frac{r_h}{\sqrt{\alpha}}.
\end{equation}
We can evaluate this cutoff across the evolutionary stages of the black hole:
\begin{itemize}
    \item At the instability onset ($r_h = r_c^{(5)} \simeq 3.108\sqrt{\alpha}$): requiring $\lambda_\ell \ge l_s$ gives $\ell_{\max}(r_c^{(5)}) \simeq 2.22 \times 3.108 \simeq 6.9$ (so only harmonics $\ell \le 6$ remain within the EFT). Because the BDG scalar instability is an eikonal effect requiring $\ell \gg 1$, its onset $\ell_{\min}$ lies safely above $\ell_{\max}(r_c^{(5)})$.
    \item At the thermodynamic turning point ($r_h = 2\sqrt{\alpha}$): as the black hole cools towards $T_H = T_{\max}$, the cutoff tightens further to $\ell_{\max} \simeq 2.22 \times 2 \simeq 4.4$ ($\ell \le 4$).
    \item Approaching the cold remnant ($r_h \to 0$): the cutoff shrinks to zero, $\ell_{\max}(r_h \to 0) \to 0$.
\end{itemize}
Thus, as the evaporating black hole enters the positive-heat-capacity branch and cools ($T_H \to 0$), the high-$\ell$ scalar modes required to destabilise the horizon are pushed outside the physical spectrum by the string UV cutoff. Provided $\ell_{\min}(r_h) > \ell_{\max}(r_h)$ holds throughout $0 < r_h \le r_c^{(5)}$, the 5D Boulware--Deser remnant is protected from the classical high-harmonic BDG instability.

Closing this section, we point out that for the relation $\ell_s=\sqrt{8\alpha}$ between
the Gauss--Bonnet coupling and the string length, Eq.~(\ref{alphastring}), the
instability region lies almost entirely outside the regime in which the truncated EGB
theory is controlled: the BDG instability requires high angular harmonics
($\ell\gg1$), whereas the derivative expansion permits only $\ell\lesssim\ell_{\max}
\simeq7$ at the onset radius. Within the controlled window the classical instability is
therefore absent. The truncated theory does not, however, establish that the instability
fails to reappear in the ultraviolet-complete theory, nor that a remnant survives there;
both questions require additional input from the UV completion and from the formation
dynamics.


\section{Observational and Experimental Signatures}
\label{sec:Experimental_Signatures}

Having established the thermodynamic phase structure, dynamical stability bounds, and the EFT domain of validity in the preceding sections, we now synthesize the cosmological, astrophysical, and collider signatures of EGB black holes in the Dark Dimension scenario. Throughout this section, the derived observables should be understood as a classical and semiclassical extrapolation of EGB gravity,
truncated at leading $\mathcal{O}(\alpha')$ higher-curvature order, into regimes that approach or probe the fundamental bulk scale $M_*$ and string length $\ell_s$. While these calculations capture the characteristic scaling laws induced by higher-curvature corrections and provide concrete phenomenological benchmarks for upcoming searches, a complete, ultraviolet-complete theory of quantum gravity, incorporating non-perturbative string dynamics, the infinite tower of higher-derivative curvature corrections, and strong gravitational back-reaction, could quantitatively modify these predictions or alter the final physical picture near the evaporation endpoint.

Within this framework, the single- and two-dark-dimension realizations point to two complementary experimental frontiers:
\begin{enumerate}
    \item \underline{Five-dimensional Dark Dimension ($D=5$, $M_* \sim 10^{10}\text{ GeV}$)} 
    \\
    Primordial black hole formation and evolution leave indirect cosmological and astrophysical imprints, comprising a scalar-induced stochastic gravitational-wave background (SGWB) in the GHz band, the absence of high-energy gamma-ray and graviton evaporation bursts, and ultra-high-frequency gravitational waves from potential remnant binary mergers.
    \item \underline{Six-dimensional Dark Dimension ($D=6$, $M_* \sim 10\text{ TeV}$)}
    \\
    Microscopic black holes accessible at future high-energy facilities such as the Future Circular Collider (FCC) exhibit distinctive kinematic benchmarks—including suppressed final-state multiplicities, a steeper energy scaling of the production cross section, and a shifted logarithmic temperature--mass slope—that sharply distinguish 6D EGB gravity from classical General Relativity.
\end{enumerate}

\subsection{Cosmological Abundance of 5D Gauss-Bonnet Dark Matter Relics}
	\label{sec:abundance}

	In this subsection, we assess the cosmological viability of 5D Einstein--Gauss--Bonnet (EGB) black hole remnants as a candidate for cold dark matter. We compute the present-day relic-density fraction $f_{\text{remnant}} \equiv \Omega_{\text{remnant}}/\Omega_{\text{DM}}$ by connecting early-universe inflation, horizon re-entry, and mass-ratio suppression to the stable EGB remnant mass scale $M_{\text{remnant}} \approx 1.06 \times 10^{-12}\,\text{g}$ (equivalently $5.92 \times 10^{11}\,\text{GeV}$). Although the stationary EGB solution does not definitively establish that such a stable remnant exists, it is nonetheless important to determine whether these putative remnants could account for the dark matter.
	
	\subsubsection{PBH Formation Formalism in Radiation Domination, Cosmological Redshift and  Mass-Ratio Suppression}
	
	Primordial black holes form during the radiation-dominated era when large-amplitude scalar curvature perturbations re-enter the cosmic horizon and undergo gravitational collapse. Assuming a Gaussian distribution of primordial overdensities $\delta$,  the initial mass fraction of the universe that collapses into PBHs of mass $M_0$ at formation time $t_f$, denoted by $\beta(M_0)$, is governed by the Press-Schechter   formula~\cite{Press:1973iz}:
	\begin{equation}
		\beta(M_0) \equiv \frac{\rho_{\text{PBH}}(t_f)}{\rho_{\text{tot}}(t_f)} = \frac{1}{\sqrt{2\pi}\sigma_R} \int_{\delta_c}^\infty \exp\left(-\frac{\delta^2}{2\sigma_R^2}\right) d\delta = \frac{1}{2} \text{erfc}\left( \frac{\delta_c}{\sqrt{2}\sigma_R} \right),
		\label{eq:beta_press_schechter}
	\end{equation}
	where $\text{erfc}(x)$ is the complementary error function which is given by: 
	$$\text{erfc}(x) \equiv 1 - \text{erf}(x) = \frac{2}{\sqrt{\pi}}\int_{x}^{\infty} e^{-t^2} \, dt~. $$
	In addition,  $\delta_c$ denotes the critical threshold for collapse during radiation domination. We should mention that the critical density threshold for collapse, $\delta_c \approx 0.45$, was evaluated strictly within the framework of four-dimensional spherical collapse for a radiation-dominated fluid with $w = 1/3$. In five dimensions, the dynamics and geometry of gravitational collapse are altered, so the higher-dimensional threshold $\delta_c^{(5\text{D})}$ departs from this standard value.
Taking this into account, we estimate the variance of density fluctuations
   by evaluating the	integral over the moving Fourier wavenumber $k$  
    \begin{equation}
\sigma_R^2 = \frac{16}{81} \int_0^\infty \frac{dk}{k}\,(kR)^4 W^2(kR)\,\mathcal{T}_{\mathrm{rad}}^2(kR)\,\mathcal{P}_{\mathcal{R}}(k).
\end{equation}
smoothed on a characteristic comoving radius $R$. The function $W(kR)$ is the Fourier transform of a spatial smoothing window and the transfer function \({\mathcal T}(k)\) describes how primordial perturbations are modified by the physical processes that occur as the Universe evolves. Finally, $\mathcal P_{\mathcal R}(k)$ is the dimensionless primordial power spectrum of the comoving curvature perturbation $\mathcal R$.
	Under the standard assumption of a narrow, delta-function-like peak in the primordial curvature power spectrum centered at the horizon scale $k_* \approx R^{-1}$, and adopting a sharp $k$-space window function $W(k_* R) \approx 1$, the integral reduces directly to the algebraic form
	\begin{equation}
		\sigma_R^2 \sim\mathcal{P}_\zeta(k_*)~.
	\end{equation}
This expresses the smoothed variance of density fluctuations in terms of the peak amplitude of the primordial curvature power spectrum $\mathcal{P}_\zeta(k_*)$ at scale $k_*$\cite{Harada:2013epa}.

	Although alternative smoothing filters (such as Gaussian or real-space top-hat window functions) or extended power spectra introduce an ${\cal O}(1)$ numerical correction to the variance prefactor, this algebraic relation still constitutes a robust and widely established benchmark across the literature for monochromatic PBH formation $k_*$\cite{Harada:2013epa}.

	Between the time the PBHs-forming scale re-enters the Hubble horizon and the
	epoch of radiation-matter equality, the energy density of the radiation
	background is diluted as $\rho_{\text{rad}}\propto a^{-4}$, whereas the energy
	density of the non-relativistic PBHs is diluted as
	$\rho_{\text{PBH}}\propto a^{-3}$. Consequently, the relative energy share of
	black holes grows linearly with the scale factor, since
	\begin{equation}
		\frac{\rho_{\text{PBH}}}{\rho_{\text{rad}}}
		\propto \frac{a^{-3}}{a^{-4}}
		= a
		\propto T^{-1}.
		\label{PBHgrowth}
	\end{equation}
	The initially small PBH fraction can thus grow relative to the radiation background until matter-radiation equality.

	As a PBH evaporates, its horizon radius $r_h$ shrinks. In 5D EGB gravity,
	higher-curvature terms make the Hawking temperature reach a maximum at
	$r_h=2\sqrt{\alpha}$ and then drop to zero as $r_h\to 0$. This stops Hawking
	radiation and leaves behind a stable remnant with residual mass
	\begin{equation}
		M_{\text{remnant}}
		=
		\frac{3\pi\alpha}{4G_5}
		\approx 1.06 \times 10^{-12}\text{ g}
		\label{Mremnant}
	\end{equation}
	Because the temperature vanishes as $r_h\to 0$, the remnant is completely
	cold and does not radiate, so it remains stable over cosmological timescales.
	
	The present abundance of these remnants depends on how much of the
	initial PBH mass is lost before the remnant forms. Taking into account this fractional
	mass loss, the present density fraction
	$f_{\text{remnant}}(M_0)$ is given by~\cite{Carr:2009jm}:
	\begin{equation}
		f_{\text{remnant}}(M_0)
		\equiv
		\frac{\Omega_{\text{remnant}}}{\Omega_{\text{DM}}}
		\approx
		2.7 \times 10^8
		\left( \frac{M_{\text{remnant}}}{M_0} \right)
		\left( \frac{M_\odot}{M_0} \right)^{1/2}
		\beta(M_0)
		\left( \frac{g_*(T_f)}{106.75} \right)^{-1/4}.
		\label{eq:f_remnant_general}
	\end{equation}
	Here, the factor $M_{\rm remnant}/M_0$ is the suppression of the mass-ratio
	which measures how much of the initial mass survives in the
	remnant. This factor is the key quantity that determines whether the remnants
	can be a significant component of dark matter.
	
	For standard macroscopic PBHs, the  suppression of the mass-ratio is enormous. If
	PBHs form with $M_0\sim 10^{15}\,$g, Hawking radiation carries away
	about $27$ orders of magnitude of the initial mass before the remnant forms,
	so 
	\begin{equation}
		\frac{M_{\text{remnant}}}{M_0}
		\sim
		\frac{10^{-12}\,\text{g}}{10^{15}\,\text{g}}
		=
		10^{-27}.
	\end{equation}
	With such a tiny mass ratio, the current remnants are negligible,
	$\Omega_{\text{remnant}}\sim 10^{-27}\,\Omega_{\text{DM}}$, regardless of the initial collapse fraction $\beta(M_0)$.  Hence, even if the initial PBH abundance were appreciable, the corresponding remnant abundance would be strongly suppressed.
	
	This observation has an important consequence. Heavy PBHs that undergo a long period of Hawking evaporation are generally inefficient sources of the present-day EGB remnant dark matter because almost all of their initial mass is radiated away. Conversely, PBHs produced with masses not far above the remnant scale,	$M_0\gtrsim M_{\text{remnant}}$, 
	suffer much less mass loss and can retain a significant fraction of their initial energy density. Therefore, the possibility of obtaining
	\begin{equation}
		f_{\rm rem}\sim 1~,
	\end{equation}
	depends not only on the primordial collapse fraction $\beta(M_0)$ but also on the location of the PBH mass distribution relative to the EGB remnant mass.
	This is the regime that we consider next.

	\subsubsection{Initial Collapse Fraction for 100$\%$ Dark Matter}

	To avoid mass-ratio suppression, the power spectrum must peak at very small scales. Then PBHs form with masses near the Gauss–Bonnet remnant scale,
	\begin{equation}
		M_0 \approx M_{\text{remnant}} \approx 1.06 \times 10^{-12}\text{ g} \quad \implies \quad \left(\frac{M_{\text{remnant}}}{M_0}\right) \approx 1.0~.
	\end{equation}
	In this regime, the PBH forms with a mass close to the remnant scale, so there is essentially no mass loss.
	Substituting $M_0 = M_{\text{remnant}} =1.06 \times 10^{-12}\text{ g}$ into 
	Eq.~\eqref{eq:f_remnant_general} yields the direct linear abundance relation:
	\begin{equation}
		f_{\text{remnant}} \approx1.18 \times 10^{31}\,\beta(M_{\text{remnant}}).
		\label{eq:f_remnant_linear}
	\end{equation}
	
	Saturating the entire dark matter content of the universe ($f_{\text{remnant}} = 1.0$) requires an initial collapse fraction of:
	\begin{equation}
		\beta_{\text{req}} = 8.51 \times 10^{-32}.
	\end{equation}
	
	Using the inverse complementary error function in Eq.~\eqref{eq:beta_press_schechter}, an initial collapse fraction of $\beta \approx8.51 \times 10^{-32}$ translates to a required fluctuation variance of $\sigma_R \approx 0.0381$ This establishes the required peak amplitude for the inflationary power spectrum:
	\begin{equation}
		\mathcal{P}_\zeta(k_*) \sim \sigma_R^2 \sim  10^{-3} \quad k_* \sim 2.1 \times 10^{28}\text{ Mpc}^{-1}
		\label{eq:power_spectrum_amplitude}
	\end{equation}
	
	The horizon re-entry wavenumber $k_* \sim  10^{28} \text{ Mpc}^{-1}$ is about $28$ orders of
	magnitude larger than the CMB pivot scale ($k_{\text{CMB}}=0.05\,\text{Mpc}^{-1}$). This power spectrum enhancement
	($\mathcal{P}_\zeta\sim 10^{-3}$) therefore lies far beyond the range probed by CMB spectral distortion and $\mu$-parameter bounds, which apply to scales $k\sim 10$-$10^4\,\text{Mpc}^{-1}$. Thus, 5D Gauss-Bonnet relics may provide a thermodynamically protected mechanism for $100\%$ cold dark matter.
	
	\subsubsection{Scalar-Induced Gravitational Wave Background}
	The scalar fluctuations that generate PBH formation inevitably couple to tensor perturbations at second order, producing a Stochastic Gravitational Wave Background~\cite{Ananda:2006af, Baumann:2007zm}.

	\begin{figure}[h]
		\centering
		\includegraphics[width=1.0\textwidth]{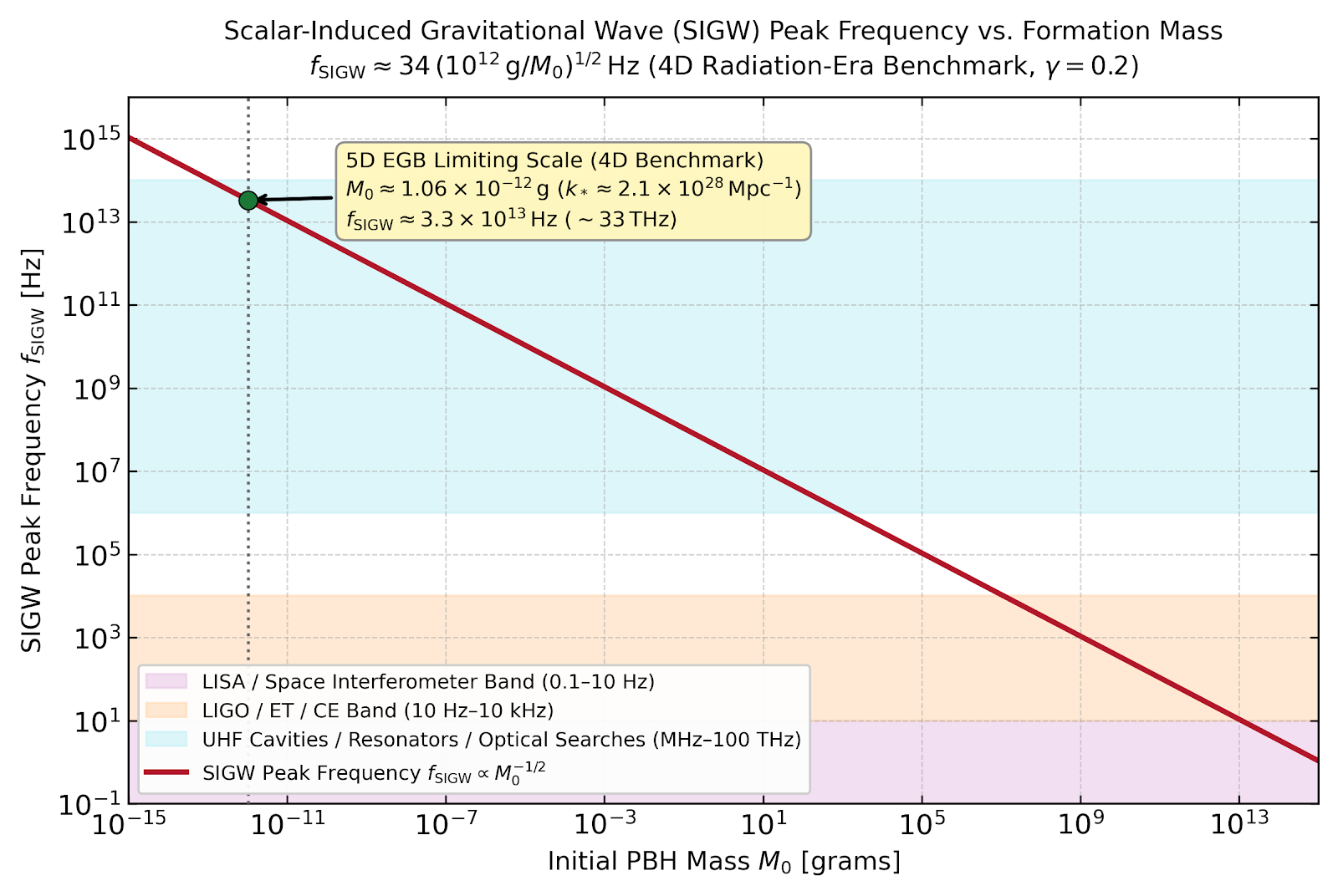}
		\caption{Peak frequency $f_{\text{SIGW}}$ of scalar-induced gravitational waves as a function of the initial PBH formation mass $M_0$.  For primordial black holes forming near the 5D EGB remnant scale $M_0 = M_{\text{remnant}} \approx 1.06 \times 10^{-12}\,\text{g}$, the resulting stochastic gravitational wave background peaks in the ultra-high frequency regime ($f_{\text{SIGW}} \approx 33\,\text{THz}$). Sensitivity bands for LISA ($\text{mHz}$), ground-based interferometers (LIGO/ET/CE, $\text{Hz}$-$\text{kHz}$), and high-frequency detectors including SRF resonators and microwave cavities ($\text{MHz}$-$\text{THz}$) are shaded for reference.}
		\label{SIGW}
	\end{figure}
For primordial black holes forming in the standard four-dimensional radiation dominated benchmark ($\gamma \approx 0.2$, $g_* = 106.75$), the comoving horizon re-entry wavenumber $k_*$ and the corresponding present-day scalar-induced gravitational-wave (SIGW) peak frequency $f_{\mathrm{SIGW}} = c k_*/(2\pi) \approx 1.55 \times 10^{-15}\,(k_*/\mathrm{Mpc}^{-1})\,\mathrm{Hz}$ scale with the formation mass $M_0$ as
\begin{equation}
k_* \approx 2.2 \times 10^{16}\,\mathrm{Mpc}^{-1}\left(\frac{M_0}{10^{12}\,\mathrm{g}}\right)^{-1/2}, \qquad
f_{\mathrm{SIGW}} \approx 34\,\mathrm{Hz}\left(\frac{M_0}{10^{12}\,\mathrm{g}}\right)^{-1/2}.
\end{equation}
Evaluating this benchmark at the 5D limiting mass scale $M_0 = M_{\mathrm{lim}} \approx 1.06 \times 10^{-12}\,\mathrm{g}$ yields a comoving peak wavenumber $k_* \approx 2.1 \times 10^{28}\,\mathrm{Mpc}^{-1}$ and a characteristic induced gravitational-wave peak in the ultra-high-frequency (UHF) regime at $f_{\mathrm{SIGW}} \approx 3.3 \times 10^{13}\,\mathrm{Hz}$ ($\sim 33\,\mathrm{THz}$), as illustrated in Figure~\ref{SIGW}. Because $k_* \approx 2.1 \times 10^{28}\,\mathrm{Mpc}^{-1}$ lies more than twenty-four orders of magnitude above the scales constrained by CMB spectral distortions ($k \sim 10\text{--}10^4\,\mathrm{Mpc}^{-1}$), the required primordial enhancement is unaffected by those bounds. Note that identifying the characteristic frequency band does not by itself establish observational detectability, which depends on the induced tensor spectral amplitude $\Omega_{\mathrm{GW}}(f)$ relative to a specific detector sensitivity curve.

    \subsubsection{Extended PBH mass functions and the remnant overabundance constraint}
\label{subsec:extended_remnant_abundance}

The preceding abundance estimate assumes, for simplicity, a nearly
monochromatic population of primordial black holes (PBHs) with initial mass
$M_0$. In particular, when the ratio $M_{\rm rem}/M_0$ is of order unity,
the final abundance of relics is not strongly suppressed by the mass lost during
the evaporation. This observation raises a natural consistency question: since
PBHs of different masses may form at different cosmological times, could a
continuous range of initial masses contribute appreciably to the remnant
density, thereby exceeding the observed dark matter abundance?

To address this question, one must distinguish the characteristic PBH mass
available at a given formation time from the probability of producing PBHs
at that time. In a standard four-dimensional, radiation-dominated epoch,
the horizon mass scales as $M_H(t_f)\propto t_f$, so that successive horizon
entry times probe different characteristic PBH masses. Nevertheless, this relation does
not imply comparable PBH formation efficiencies at all
masses because the latter depend sensitively on the scale dependence of the
primordial perturbations. If  $\beta_{\ln{M}}(M_0)$ denotes the fraction of the total energy density
converted into PBHs at formation per logarithmic interval of their
initial mass, then $\beta_{\ln{M}}(M_0)\,d\ln{M_0}$ is the contribution of
that interval to the initial fraction of the PBH energy. In the limit in which
PBHs remain a subdominant component until evaporation and their remnants
subsequently behave as nonrelativistic matter, their present-day density
fraction can be written schematically as
\begin{equation}
	\label{eq:extended_remnant_fraction}
	f_{\rm rem}^{\rm tot}
	\equiv \frac{\Omega_{\rm rem}}{\Omega_{\rm DM}}
	= \int_{\ln M_{\min}}^{\ln M_{\max}} \mathcal{A}(M_0)\,
	\frac{M_{\rm rem}(M_0)}{M_0}\,
	\beta_{\ln M}(M_0)\,d\ln M_0\,.
\end{equation}
Here $\mathcal{A}(M_0)$ is the cosmological conversion factor from the
initial PBH energy fraction to the present-day dark matter fraction,
including the relative redshifting of matter and radiation and the
appropriate changes in relativistic degrees of freedom. In equation
\eqref{eq:extended_remnant_fraction} it is assumed that $\beta_{\ln M}$ and
$\mathcal{A}$ are defined consistently for each formation epoch. For simplicity, we assume a
standard radiation-dominated history, where the factor $\mathcal{A}$ increases for earlier formation times. 

For the 5D EGB branch considered here, the limiting mass $M_{\rm rem}$ is set by the Gauss--Bonnet coupling and the gravitational parameters. For fixed model parameters, it is therefore natural to regard $M_{\rm rem}$ as approximately independent of $M_0$, provided the assumed endpoint is reached. The mass-weighting factor in \eqref{eq:extended_remnant_fraction} then approaches unity for
$M_0\gtrsim M_{\rm rem}$, whereas it suppresses contributions from
$M_0\gg M_{\rm rem}$. Consequently, PBHs formed with masses close to the
limiting mass can be particularly important. Nevertheless, the mere
existence of several such mass intervals does not imply overproduction:
their abundances are determined by the same underlying primordial power
spectrum and cannot be normalized independently.

For Gaussian density perturbations, a widely used illustrative estimate for the collapse 
fraction is given by Eq.~(\ref{eq:beta_press_schechter}),
\begin{equation}
	\label{eq:beta_gaussian_extended}
	\beta_{\rm coll}(M_0) \simeq
	\frac{1}{2}\,\operatorname{erfc}\!\left(
	\frac{\delta_c}{\sqrt{2}\,\sigma(M_0)}\right)
	\simeq
	\frac{\sigma(M_0)}{\sqrt{2\pi}\,\delta_c}
	\exp\!\left[-\frac{\delta_c^2}{2\sigma^2(M_0)}\right],
	\qquad \sigma(M_0)\ll\delta_c\,.
\end{equation}

Observe that equation~\eqref{eq:beta_gaussian_extended}  illustrates the crucial exponential sensitivity to the variance $\sigma(M_0)$.
A relatively modest reduction of the variance away from its maximum can suppress PBH formation by many orders of magnitude.

For  a localized enhancement of the primordial curvature power spectrum may be parameterized as
\begin{equation}
	\label{eq:peaked_power_extended}
	\mathcal{P}_{\mathcal R}(k)
	= \mathcal{P}_{\rm bg}(k)
	+ A_{\mathcal R}\exp\!\left[
	-\frac{\ln^2(k/k_*)}{2\Delta^2}\right],
\end{equation}
where $k_*$ determines the characteristic formation scale and $\Delta$
controls the logarithmic width of the enhancement. After convolution
with the appropriate window and transfer functions, such a spectrum can
produce a PBH mass distribution concentrated around a preferred mass
range. PBH formation outside that range may be exponentially suppressed,
so that the integral in Eq.~\eqref{eq:extended_remnant_fraction} remains
finite and compatible with the observed dark matter abundance even when
$M_{\rm rem}/M_0=\mathcal{O}(1)$ near its dominant contribution. 
The necessary abundance constraint is therefore
\begin{equation}
	\label{eq:remnant_no_overclosure}
	\displaystyle
		\int \mathcal{A}(M_0)\,
		\frac{M_{\rm rem}(M_0)}{M_0}\,
		\beta_{\ln M}(M_0)
		\;d\ln M_0\;\leq\;1\,,
\end{equation}
with approximate equality if the remnants constitute all of the dark
matter. Conversely, if the contribution per logarithmic mass interval
were of order unity over several independent intervals, their sum would
indeed overproduce dark matter. Therefore, a monochromatic estimate that
yields $f_{\rm rem}\simeq1$ should be interpreted as a reference for
a localized mass distribution, not as a normalization that can be
applied separately to every possible PBH mass.

Finally we wish to emphasize that, implementing Eq.~\eqref{eq:remnant_no_overclosure} in the Dark Dimension scenario is not straightforward. 
The relations $M_H\propto t_f$, the usual radiation-era redshifting factor, 
and standard estimates of the collapse threshold are four-dimensional benchmarks. 
If PBH formation occurs during a genuinely higher-dimensional cosmological phase,
the mass–time relation, formation criterion, and cosmological conversion factor must 
be derived for that background. Moreover, treating the zero-temperature EGB endpoint 
as a surviving remnant remains an assumption of the effective description. Subject to these qualifications, 
the absence of overclosure constrains not only the amplitude of the primordial enhancement but also its width 
and position, providing an additional consistency condition on the remnant dark matter scenario.

	     \subsubsection{Observational Imprints of 5D EGB PBHs from Gravitational Waves}

	     EGB gravity in the Dark Dimension scenario changes black hole thermodynamics and horizon geometry. This leaves three main types of signals in the gravitational wave sky: secondary gravitational waves from the early universe, modified ringdown spectra, and ultra-high-frequency waves from PBH mergers and stable remnants.

	\subsubsection{Suppression of High-Energy Evaporation Bursts}
	
	In standard General Relativity, the final explosion of an evaporating black hole produces a sudden, high-energy burst of gravitons and Standard Model particles. In a $D$-dimensional bulk spacetime, this final phase produces an ultra-high-frequency gravitational wave signal.
	
	In contrast, the 5D EGB metric function $f(r)$ forces the Hawking temperature $T_H(r_h)$ to vanish as $r_h \to 0$, as can be inferred from Eq. (11). Consequently, the high-frequency graviton burst characteristic of the $T \to \infty$ endpoint in standard GR is completely absent. As $r_h \rightarrow 2\sqrt{a}$, the emission of tensor modes drops smoothly to zero, leaving a quiet, non-radiating remnant.
	The absence of the high-energy gamma-ray and graviton bursts expected from evaporating sub-solar PBHs agrees with the EGB thermodynamic cutoff.

	\subsubsection{Modified Quasi-Normal Modes and Remnant Mergers}
	
The presence of higher-curvature Gauss-Bonnet terms introduces important differences between EGB black holes and their counterparts in classical General Relativity. In particular, the quasi-normal-mode (QNM) spectrum, $\omega=\omega_R-i\omega_I$, is modified by the Gauss-Bonnet coupling and depends on the dimensionless ratio $\alpha/r_h^2$. In the perturbative regime $\alpha/r_h^2\ll1$, the QNM frequencies can schematically be written as
\begin{equation}
\omega_{\rm EGB}\simeq\omega_{\rm GR}
\left[1+\mathcal{C}_D\frac{\alpha}{r_h^2}+\cdots\right],
\end{equation}
where $\mathcal{C}_D$ depends on the dimension of spacetime and the perturbation mode. For the near-remnant regime considered here, however, $\alpha/r_h^2$ is of order unity, and the full EGB perturbation spectrum is therefore required rather than a perturbative expansion around GR.

As discussed in Section 3, five-dimensional EGB primordial black holes can avoid complete evaporation. Instead, their Hawking temperature decreases towards zero, leaving stable, cold remnants with mass
$M_{\rm rem}\simeq 1.06 \times 10^{-12}\text{ g}$. In principle, such remnants could form bound pairs in the early Universe or in sufficiently dense dark-matter environments. If a binary forms, its orbital dynamics could lead to the emission of gravitational radiation and, eventually, to a merger. For a compact object of characteristic size $r_h$, the corresponding dynamical frequency is parametrically of order
\begin{equation}
f_{\rm char}\sim\frac{c}{2\pi r_h}.
\end{equation}

For the EGB remnants considered here, $r_h\sim\sqrt{\alpha}\sim10^{-26}\,\mathrm{m}$, giving a characteristic frequency of order
\begin{equation}
f_{\rm char}\sim
\frac{c}{2\pi\sqrt{\alpha}}
\sim {\cal O}(M_*),
\end{equation}
where we have restored the appropriate conversion between natural and SI units. For $M_*\sim10^{10},\mathrm{GeV}$, this corresponds parametrically to frequencies of order $10^{33}\,\mathrm{Hz}$.

This estimate should be regarded as a characteristic dynamical scale rather than a prediction for an observable merger frequency. At such extremely small masses and length scales, the Compton wavelength of the remnant $\lambda_C\sim \hbar/(M c)\approx 3.3 \times 10^{-28}\text{ m}$, is comparable to its horizon size, indicating that quantum-gravitational effects may become important. Moreover, the propagation of gravitational perturbations in the five-dimensional bulk and their projection onto the effectively four-dimensional observable sector require a dedicated analysis. Consequently, whether binary remnants can produce an observable ultra-high-frequency gravitational-wave signal remains an open question.

\begin{figure}[h]
		\centering
	\includegraphics[width=1.0\textwidth]{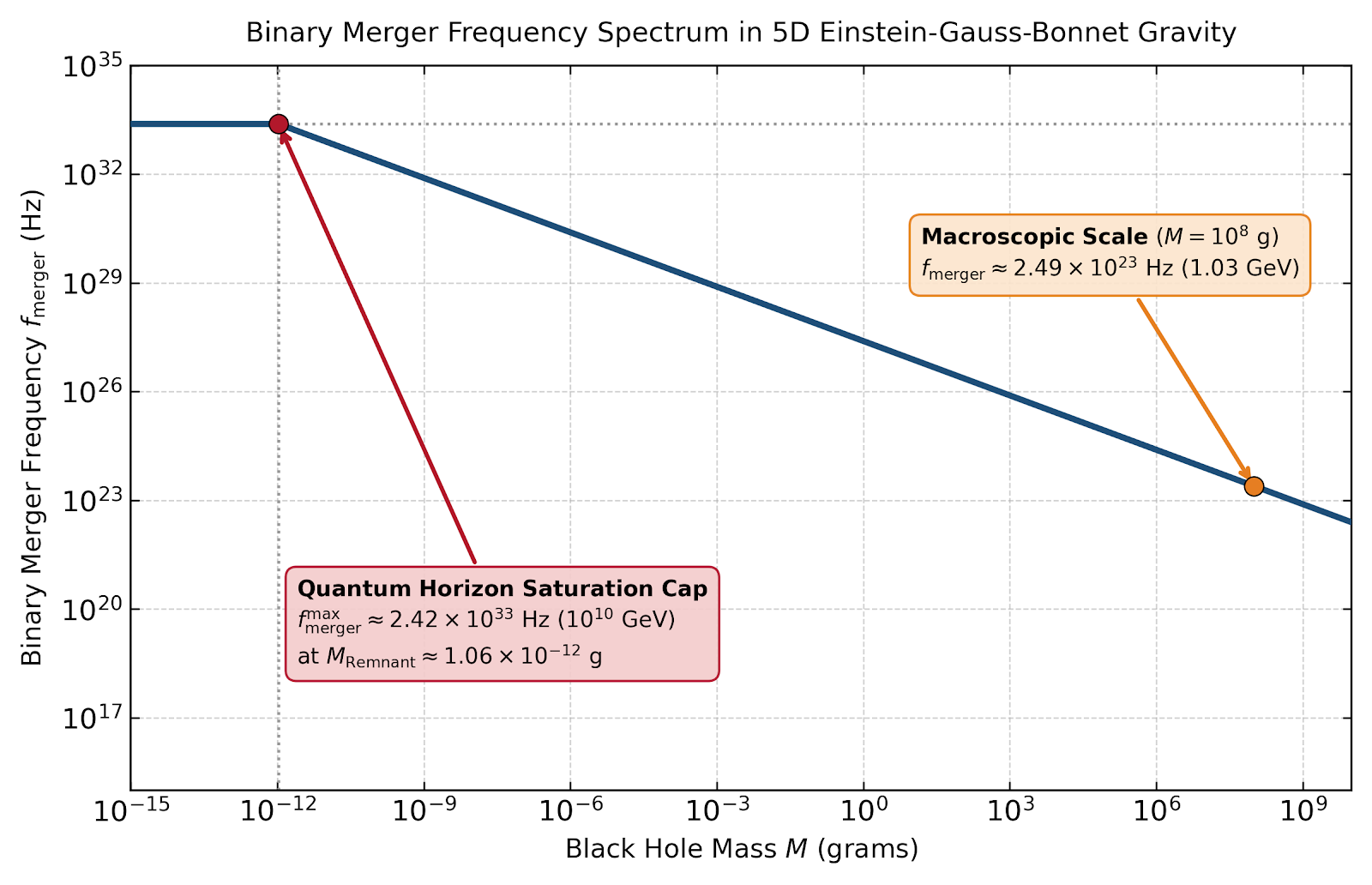}
		\caption{Binary Merger frequency as a function  of black hole mass}
		\label{Merger}
	\end{figure}

To illustrate the connection between black hole mass and merger frequency, we show the merger frequency as a function 
of black hole mass in Fig.~\ref{Merger}. Ground-based interferometers such as LIGO operate in the Hz–kHz band and 
therefore are unable to detect UHFGWs. Future non-interferometric UHF detectors may nevertheless push the sensitivity 
of gravitational-wave searches to much higher frequencies.

 \subsection{Experimental Signatures of 6D EGB Micro Black Holes at Future Colliders}

High-energy collisions at future colliders, such as the proposed FCC operating at a center-of-mass energy of $\sqrt{s} = 100\,\text{TeV}$, offer a unique window into the two dark dimensions scenario ($D=6$, $n=2$), where the fundamental gravity scale is lowered to $M_* \approx 10\,\text{TeV}$. When the collision energy exceeds the species scale ($E \ge M_*$), microscopic black holes can be produced with a cross-section scaling with the horizon area, $\sigma_{\text{prod}} \propto \pi r_h^2$.

The evaporation of 6D EGB micro black holes yields dramatic experimental signatures that fundamentally differ from both standard 6D General Relativity (GR) and pure Standard Model backgrounds across three primary observables:

\subsubsection{ Suppressed Multiplicity and Hard Particle Emission.}
In standard 6D General Relativity ($\alpha = 0$), the Hawking temperature scales inversely with the horizon radius as:
\begin{align}
    T_H^{\text{GR}} = \frac{3}{4\pi r_{\text{GR}}} = \left(\frac{9}{4\pi}\right)^{1/3}M_*\left(\frac{M_{\text{BH}}}{M_*}\right)^{-1/3},
\end{align}
yielding a average decay multiplicity of \cite{PKanti}:
\begin{equation}
   \langle N \rangle_{\text{GR}} \approx \frac{M_{\text{BH}}}{2T_H^{\text{GR}}} = \left(\frac{\pi}{18}\right)^{1/3}\left(\frac{M_{\text{BH}}}{M_*}\right)^{4/3}
\end{equation}
For a $100\,\text{TeV}$ black hole at $M_* = 10\,\text{TeV}$, classical 6D GR predicts a high-multiplicity thermal explosion into $\langle N \rangle_{\text{GR}} \approx 12$ secondary particles \cite{Leontaris:2026kvu}.

In contrast, for microscopic 6D EGB black holes ($r_h \ll \sqrt{\alpha}$), the linear mass-horizon relation $M_{\text{BH}} \approx 32\pi^2 M_*^2 r_h$ drives the Hawking temperature up to:
\begin{equation}
    T_H^{\text{EGB}} \approx \frac{1}{8\pi r_h} \approx \frac{4\pi M_*^2}{M_{\text{BH}}}
\end{equation}

 Consequently, the average particle multiplicity in 6D EGB gravity is suppressed to:
\begin{equation}
   \langle N \rangle_{\text{EGB}} \approx \frac{M_{\text{BH}}}{2T_H^{\text{EGB}}} \approx \frac{1}{8\pi}\left(\frac{M_{\text{BH}}}{M_*}\right)^2
\end{equation}
In Table~\ref{Table2} we summerize the thermodynamic and collider observables for 6D micro black holes between EGB gravity and GR gravity with energy scale close to $M_*$.
For $M_{\text{BH}} = 100\,\text{TeV}$, the 6D EGB Hawking temperature rises to $T_H^{\text{EGB}} \approx 12.6\,\text{TeV}$, drastically suppressing the particle multiplicity down to $\langle N \rangle_{\text{EGB}} \approx 4$. Instead of a spherical ``fireball" of many soft particles, a 6D EGB micro black hole decays instantaneously into a small number of extremely energetic particles. This example suggests that the semiclassical approximation may break down, requiring a full quantum gravity treatment to explore this energy regime. 
It is unclear whether the semiclassical approximation remains valid for  $M_{BH}\gtrsim $ 200 TeV; consequently, our calculation cannot definitively exclude remnant formation, as unknown effects may intervene.

\begin{table}[t]
\centering
\caption{Stationary EGB thermodynamic quantities, illustrative geometric and multiplicity estimates of 6D EGB microscopic black holes across benchmark masses ($D=6$, $M_* = 10\text{ TeV}$, $G_6 = (8\pi M_*^4)^{-1}$, $\alpha = M_*^{-2}$).}
\label{tab:6D_benchmarks}
\begin{tabular}{ccccccccc}
\hline\hline
$M$ & $M/M_*$ & $r_h$ & $T_H$ & $\langle E \rangle \approx 3T_H$ & $\langle E \rangle / M$ & $S_{\text{Wald}}$ & $\langle N \rangle$ & $\sigma \approx \pi r_h^2$ \\
$[\text{TeV}]$ & & $[10^{-21}\text{ m}]$ & $[\text{TeV}]$ & $[\text{TeV}]$ & & & & $[\text{pb}]$ \\
\hline
$100$ & $10$ & $0.62$ & $12.57$ & $37.7$ & $37.7\%$ & $3.98$  & $\sim 4$    & $0.0123$ \\
$150$ & $15$ & $0.94$ & $8.39$  & $25.2$ & $16.8\%$ & $8.95$  & $\sim 9$    & $0.0276$ \\
$200$ & $20$ & $1.25$ & $6.30$  & $18.9$ & $9.4\%$  & $15.90$ & $\sim 16$   & $0.0490$ \\
$500$ & $50$ & $3.11$ & $2.55$  & $7.6$  & $1.5\%$  & $98.75$ & $\sim 98$   & $0.3041$ \\
\hline\hline
\end{tabular}\label{Table2}
\end{table}

\subsubsection{Temperature-Mass Slope Shift in the $\log T_H - \log M_{\text{BH}}$ plane.}
	The energy threshold for black hole production at colliders is determined by the condition $\lambda_C < r_h$, 
    where $\lambda_C$ is the Compton wavelength,
	\begin{equation}
		\lambda_C = \frac{4\pi}{E} < \frac{1}{\sqrt{\pi} M_*} \left(\frac{E}{M_*}\right)^{\frac{1}{\mathfrak{n}+1}} \left( \frac{8\,\Gamma\!\left(\frac{\mathfrak{n}+3}{2}\right)}{\mathfrak{n}+2} \right)^{\frac{1}{\mathfrak{n}+1}},
	\end{equation}
	with $E$ being the total energy of the colliding partons that form the black hole.
	
	The production cross-section in the center-of-mass energy of the collision in classical general relativity is proportional to the horizon area and depends with energy is given by  \cite{PKanti}:
    \begin{equation}
         \sigma_{\text{prod}}^{GR} \propto \pi r_h^2 \approx \pi \left( \frac{3}{16\pi^2} \right)^{2/3} \frac{1}{M_*^2} \left( \frac{E}{M_*} \right)^{2/3}.
    \end{equation}
    On the other hand the production cross section in 6D EGB is given by:
	\begin{equation}
		\sigma_{\text{prod}}^{\text{EGB}}(E) \approx \pi \left( \frac{E}{32\pi^2 M_*^2} \right)^2 = \frac{1}{1024 \pi^3 M_*^2} \left( \frac{E}{M_*} \right)^2 .
	\end{equation}

    \begin{figure}[h]
    \centering
    \includegraphics[width=0.95\textwidth]{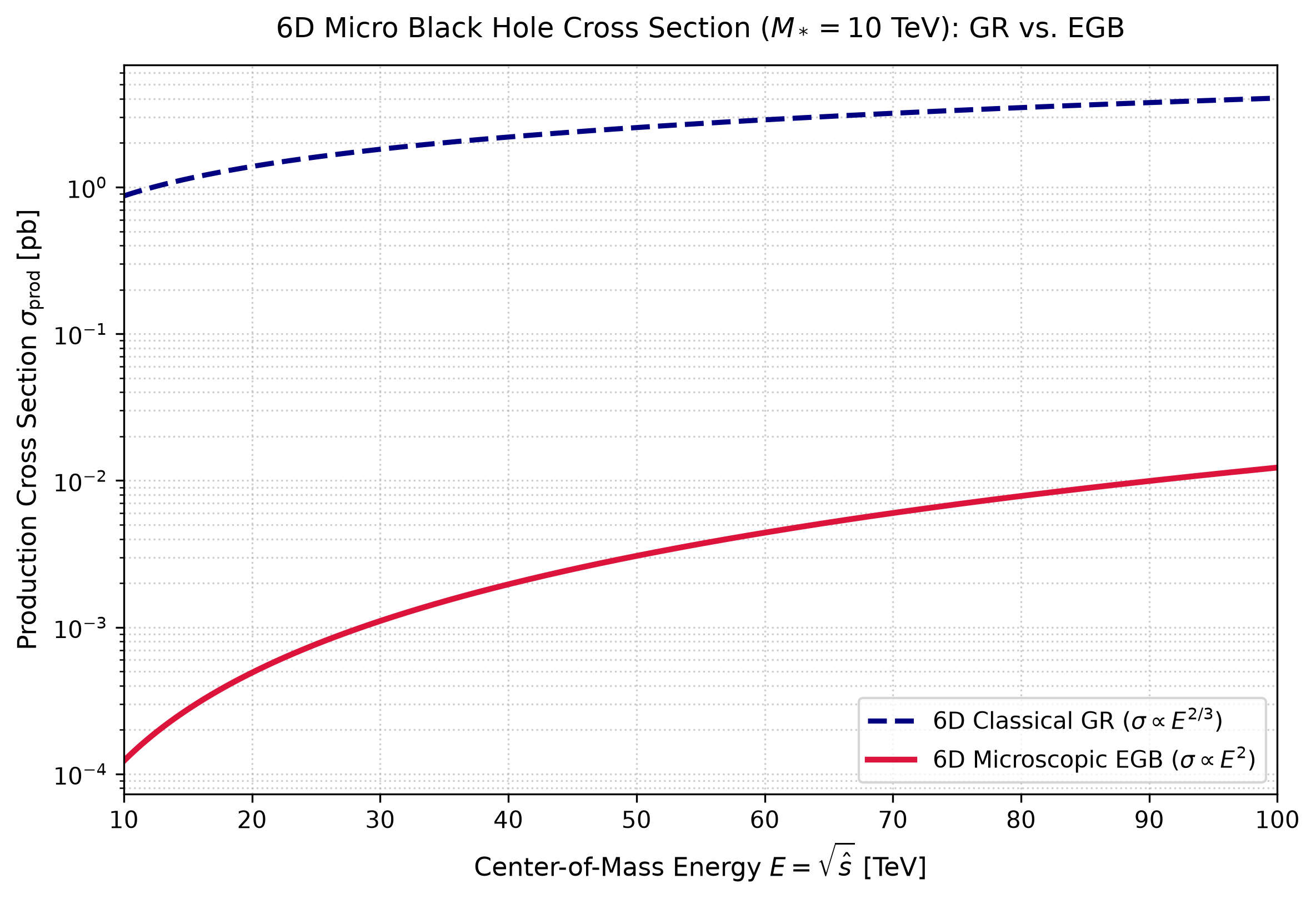}
    \caption{The production cross-section $\sigma_{\text{prod}}$ of 6D micro black holes as a function of the parton center-of-mass energy $E = \sqrt{\hat{s}}$ for a fundamental scale $M_* = 10\,\text{TeV}$, comparing classical 6D General Relativity ($\alpha = 0$, blue dashed line) with 6D EGB gravity ($\alpha = 1/M_*^2$, red solid line) on a logarithmic scale. Higher-curvature Gauss-Bonnet corrections induce a significant geometric suppression of the horizon radius at microscopic scales ($r_h \ll \sqrt{\alpha}$), reducing the cross-section magnitude while shifting its energy dependence from the shallower $\sigma_{\text{GR}} \propto E^{2/3}$ scaling to a steeper $\sigma_{\text{EGB}} \propto E^2$ power law.}
    \label{CrossSection}
\end{figure}

     At microscopic scales ($r_h \ll \sqrt{\alpha}$), higher-curvature Gauss-Bonnet terms alter the  geometry near the horizon, critically contracting the horizon radius relative to pure GR ($r_h^{\text{EGB}} \ll r_h^{\text{GR}}$). Since the geometric cross-section scales quadratically with the horizon size ($\sigma_{\text{prod}} = \pi r_h^2$), this compression suppresses the production cross-section by two to three orders of magnitude across the FCC energy domain. This can be seen in Figure~\ref{CrossSection} were we plot the production cross-section of 6D micro black holes as a function
of the parton center-of-mass energy.
   Despite the overall magnitude suppression, the microscopic 6D EGB regime exhibits a much steeper energy growth. Whereas classical 6D general relativity predicts a shallower power law,  $\sigma_{\text{GR}} \propto E^{2/3}$ , in the Gauss–Bonnet-dominated limit the linear mass–horizon relation
   ($E \approx 32\pi^2 M_*^2 r_h$), leads to an EGB cross-section that scales quadratically with energy, $\sigma_{\text{EGB}} \propto E^2$. 
This relation reveals an enhancement of the cross-section with $E$  that is absent in any Standard Model or SUSY process, and thus constitutes a distinctive signature.

	To extract useful information about  the scenario with  two dark dimensions, we must perform accurate measurements of  the mass and the temperature 
    of the produced mini black holes. In a high-energy collision of composite particles, it is not known which pair of partons led to the creation of the black hole nor what its total energy was. Therefore, the black hole mass must   be reconstructed, through the measurement of the energy of the particles appearing in the final state after evaporation.
	The temperature of the black hole can be determined by fitting the observed Hawking radiation spectrum. Particularly clean spectra can be obtained from events containing only photons and electrons in the final state, which offer very low backgrounds and excellent energy resolution even at high energies, while classical $D$-dimensional GR predicts a logarithmic slope of,
\begin{equation}
    \log(T_H^{\text{GR}}) = -\frac{1}{n+1} \log(M_{\text{BH}}) + C \qquad \xrightarrow{n=2} \qquad \text{slope} = -\frac{1}{3}~\cdot 
\end{equation}
The microscopic 6D EGB regime shifts this relation to $T_H^{\text{EGB}} \propto M_{\text{BH}}^{-1}$, resulting in a significantly steeper logarithmic slope,
\begin{equation}
    \log(T_H^{\text{EGB}}) = -\log(M_{\text{BH}}) + \log(4\pi M_*^2) \qquad \implies \qquad \text{slope} = -1
\end{equation}
Measuring a logarithmic slope of $-1$ instead of $-1/3$ (6D GR) in the $\log T_H - \log M_{\text{BH}}$ plane (see Figure~\ref{5D6Dcompare}) would indicate the presence of 6D EGB quantum gravity at future colliders.

It is important to emphasize that the thermodynamic properties of a mini-black hole in the energy scale of FCC ($\sim100\text{TeV}$) presented above,
 most notably the nominal Hawking temperature $T_H \approx 12.6\,\text{TeV}$ for an initial mass $M = 100\,\text{TeV}$, 
 are evaluated within the standard semiclassical Hawking approximation and should be regarded as an effective phenomenological 
  benchmark rather than an exact description. 
 In this regime, the black hole mass exceeds the higher-dimensional fundamental scale by only an order of magnitude ($M / M_* \sim 10$ for $M_* \sim 10\,\text{TeV}$), placing the system at the boundary between a semiclassical state and a quantum black hole (QBH). Consequently, the characteristic energy of an individual emitted quantum, $\langle E \rangle \sim 3 T_H \approx 38\,\text{TeV}$, represents a substantial fraction of the remaining black hole mass ($3 T_H / M \sim 0.38$). In this threshold domain, gravitational back-reaction cannot be neglected, and the continuous thermal evaporation picture gives way to a discrete, few-body decay into a small multiplicity of hard Standard Model quanta. Whereas  a precise determination of the decay channels and branching ratios requires a full underlying theory of quantum gravity, the classical geometric production cross section $\sigma(pp \to \text{BH}) \sim \pi r_h^2$ remains robust against ultraviolet completions, ensuring that our order-of-magnitude estimates reliably delineate the discovery potential of 6D microscopic black holes at the future colliders.

 \begin{figure}[h]
		\centering
	\includegraphics[width=0.45\textwidth]{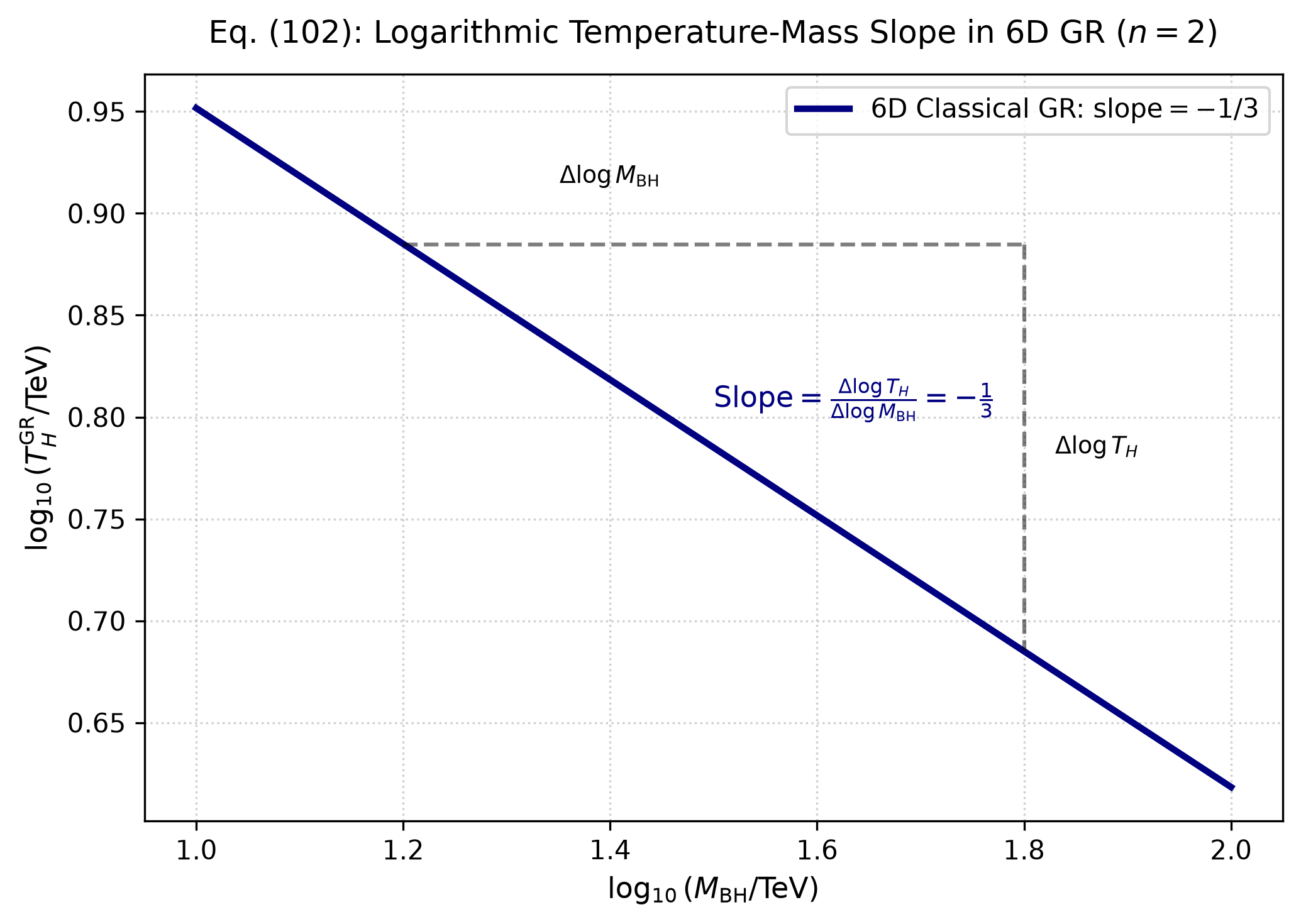}
    \includegraphics[width=0.45\textwidth]{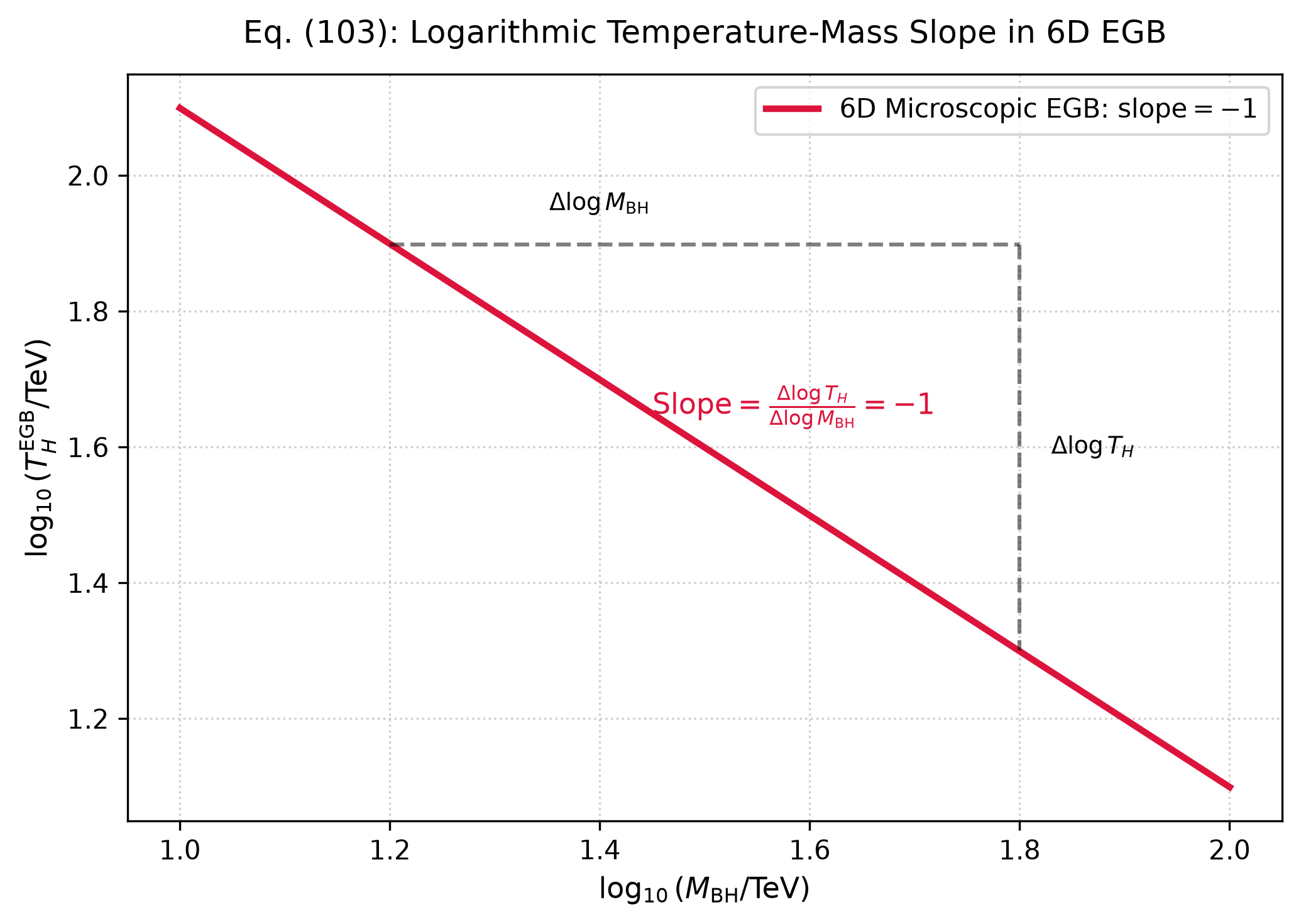}
 		\caption{Logarithmic Hawking temperature versus black hole mass ($\log_{10} T_H$ vs. $\log_{10} M_{\text{BH}}$) for 6D micro black holes ($D=6$, $n=2$, $M_* = 10\,\text{TeV}$). \textbf{Left panel:} Classical 6D General Relativity ($\alpha = 0$), exhibiting the standard logarithmic slope of $d\log T_H / d\log M_{\text{BH}} = -1/(n+1) = -1/3$. \textbf{Right panel:} Microscopic 6D EGB gravity ($\alpha = 1/M_*^2$) in the regime $r_h \ll \sqrt{\alpha}$, where higher-curvature corrections shift the mass-horizon relation to $M_{\text{BH}} \approx 32\pi^2 M_*^2 r_h$. This results in a significantly steeper logarithmic slope of $d\log T_H / d\log M_{\text{BH}} = -1$. Measuring this sharp slope shift in reconstructed collider events provides a signaturefor 6D EGB gravity, subject to formation and reconstruction uncertainties.}
		\label{5D6Dcompare}
	\end{figure}

	\section{Conclusions}
	\label{sec:Conclusions}

In this paper we have studied Einstein-Gauss-Bonnet  black holes in
the Dark Dimension scenario, with one or two extra spatial dimensions
decompactified at the micron scale. Our main results are summarized as
follows.
We first investigated the General $D$-dimensional EGB thermodynamics of PBHs 
and derived the mass, Hawking temperature, Wald entropy, and heat capacity of the
$D$-dimensional Boulware-Deser black hole.
We further showed that a stable heat capacity branch exists only in five dimensions, whilst a consequence of this fact is that the formal thermodynamic evolution can approach a cold remnant only in five dimensions 
  For more than five dimensions, $D\ge 6$, the heat capacity $C$ is negative, for all $r>0$.  Moreover, 
  we found that in $D=5$ the  temperature vanishes as $r_h\to 0$, whereas for $D\ge 6$ it diverges as
    $1/r_h$, just as in four-dimensional Schwarzschild case. The Gauss-Bonnet term
    therefore does not stabilise the black hole in $D\ge 6$; it merely reduces
    the near-horizon geometry to an effectively four-dimensional one.

In the five-dimensional Dark Dimension scenario, with $M_* \sim 10^{10}$~GeV, the Hawking temperature reaches a maximum $T_{\max} = 1/(8\pi\sqrt{\alpha})$ at $r_h = 2\sqrt{\alpha}$ and then decreases to zero, leaving an apparently remnant of mass $M_{\rm remnant} \simeq1.06 \times 10^{-12}\text{ g}$. Because the temperature vanishes as $r_h \to 0$, the remnant is completely cold.
 Whether this endpoint represents a physically stable remnant requires a treatment of the regime in which the semiclassical approximation breaks down. The transition at $r_h = 2\sqrt{\alpha}$ is a second-order phase transition, where the heat capacity diverges.
The five-dimensional Einstein-Gauss-Bonnet solution admits a positive-heat-capacity branch whose formal zero-radius limit has finite mass and vanishing Hawking temperature. However, our analysis of the dynamical stability bounds demonstrates that this entire branch lies within the classical scalar-instability region identified by Beroiz, Dotti, and Gleiser~\cite{Beroiz:2007gp}.  
Evaluating the physical wavelengths of these unstable modes on the horizon reveals that as the black hole mass approaches the formal limiting remnant mass $M_{\mathrm{lim}}$, both the horizon radius and the unstable perturbations shrink to sub-stringy length scales ($\lambda_\ell, r_h \ll l_s$). In this regime, the background derivative expansion breaks down and the perturbation spectrum can no longer be trusted without an ultraviolet-complete theory. Consequently, the truncated EGB effective field theory can definitively establish neither the persistence of the classical instability nor the stability of a surviving remnant; whether this endpoint resolves into a stable stringy or quantum remnant remains undecided and requires additional dynamical input from a full theory of quantum gravity.

Assuming that such a stable remnant does form at the mass scale $M_{\mathrm{remnant}} = M_{\mathrm{lim}} \approx 1.06 \times 10^{-12}~\text{g}$, its present-day abundance is controlled by the mass-ratio suppression $M_{\mathrm{remnant}}/M_0$. For standard PBHs with $M_0 \sim 10^{15}$~g, the suppression is $\sim 10^{-27}$, so the remnants are negligible regardless of the initial collapse fraction. Avoiding this suppression requires PBHs formed with masses close to the remnant scale $M_0 \sim M_{\rm remnant}$. In this unsuppressed regime, requiring that the remnants account for all of the dark matter fixes the initial collapse fraction to $\beta \simeq8.51 \times 10^{-32}$, corresponding to a peak curvature amplitude $\mathcal{P}_{\mathcal{R}} \simeq 10^{-3}$ at $k_* \simeq 10^{28}$~Mpc$^{-1}$. This scale lies far beyond the range probed by CMB spectral distortions and $\mu$-parameter bounds, so the corresponding enhancement of the inflationary power spectrum is not directly constrained by those observations. The scenario is therefore cosmologically viable, provided that the power spectrum peaks at the appropriate scale.

We next examine the observational imprints of the 5D EGB scenario, which 
leaves three potentially observable signatures in gravitational waves:
    \begin{itemize}
        \item a scalar-induced gravitational-wave background at THz
        frequencies, associated with the formation of the PBHs;
        \item a suppression of the high-energy gamma-ray and graviton bursts
        expected from evaporating sub-solar PBHs, consistent with the EGB
        thermodynamic cutoff;
        \item ultra-high-frequency gravitational waves from mergers of
        remnant binaries, with characteristic frequencies of order  $f\sim c/(2\pi\sqrt{\alpha})\sim 10^{33}$~Hz, 
        far beyond the reach of
        current and planned interferometers.
    \end{itemize}
    The first two signatures are within the reach of future high-frequency
    gravitational-wave detectors and of existing gamma-ray observations. The
    third is at present only a theoretical possibility, and its observability
    depends on whether remnant binaries can form and merge efficiently.

In our subsequent analysis we  studied a 6D Dark Dimension scenario with two micron-size dimension 
and a fundamental scale comparable to $M_*\sim 10$~TeV. 
In this case   the heat capacity is negative for all $r_h>0$, and the temperature
    diverges as $r_h\to 0$, hence, there is no stable remnant. Nevertheless, the Gauss-Bonnet corrections
    substantially modify the evaporation history. For macroscopic black
    holes, the lifetime is shifted by a factor
    $1+\frac{280}{3}\alpha/r_{\rm GR}^2$ relative to general relativity. For
    microscopic black holes, the lifetime scales as
    $\tau_{\rm EGB}\propto M_0^3$, much steeper than the classical
    $\tau_{\rm GR}\propto M_0^{5/3}$, and is enhanced by nearly two orders of
    magnitude at $M_0\sim 100$~TeV.

  The spectrum of  future colliders would exhibit sharply different
    signatures from their general-relativistic counterparts: a suppressed
    multiplicity ($\langle N\rangle_{\rm EGB}\sim 4$ instead of
    $\langle N\rangle_{\rm GR}\sim 12$ for $M_{\rm BH}=100$~TeV). Also, we have steeper production cross-section
    scaling $\sigma\propto E^2$ instead of $E^{2/3}$, and a logarithmic
 temperature--mass slope of $-1$ instead of $-1/3$. These features
    provide a distinctive experimental handle on six-dimensional  Gauss-Bonnet gravity.

    
Several open questions remain. The most important is the reliability of the
semiclassical approximation near $r_h\sim\sqrt{\alpha}$, where the horizon
radius is of the order of the fundamental Planck or string length. A full
quantum-gravitational treatment is needed to determine the endpoint of
evaporation in $D\ge 6$, and to confirm the stability of the $D=5$ remnant.
A second question is the formation of PBHs with masses near the fundamental
scale: it is not obvious that the standard Press-Schechter formalism applies
in this regime, and a dedicated analysis of collapse at
$r_h\sim\sqrt{\alpha}$ would be valuable. A third question concerns the
observability of the ultra-high-frequency gravitational-wave signal from
remnant mergers, which depends on the formation and merger rates of remnant
binaries in the early universe and in dense dark-matter environments. We hope
to address some of these questions in future work.
	
	\newpage
	
	\begin{center}
		{\bf\huge Appendix}
			\end{center}

\subsection{Derivation of the Wald Entropy in 5D Einstein-Gauss-Bonnet Gravity}
\label{app:wald_entropy_5D}

In this appendix, we derive the Wald entropy in five-dimensional
Einstein-Gauss-Bonnet  gravity for a static, spherically symmetric
black hole with a spherical horizon. Throughout, the Gauss--Bonnet coupling
$\alpha$ and Newton's constant $G_5$ are taken to be fixed.
The entropy is obtained directly from the Noether-charge prescription, and the conventional first law $dM=T_H\,dS$  then provides an independent consistency check for variations at fixed $\alpha$.

The action is
\begin{equation}
I=\frac{1}{16\pi G_5}\int d^5x\,\sqrt{-g}\,\mathcal L,
\qquad
\mathcal L=R+\alpha\mathcal L_{\rm GB},
\label{eq:app_action}
\end{equation}
where $\alpha$ has dimensions of length squared and
\begin{equation}
\mathcal L_{\rm GB}
=R^2-4R_{\mu\nu}R^{\mu\nu}
+R_{\mu\nu\rho\sigma}R^{\mu\nu\rho\sigma}.
\end{equation}
The scalar Lagrangian $\mathcal L$ in Eq.~\eqref{eq:app_action} does not
include the overall factor $1/(16\pi G_5)$ or the volume element $\sqrt{-g}$.

We consider the line element
\begin{equation}
ds^2=-f(r)\,dt^2+\frac{dr^2}{f(r)}+r^2d\Omega_3^2,
\end{equation}
where $d\Omega_3^2$ is the metric on the unit three-sphere, whose volume
is $\Omega_3=2\pi^2$. The horizon is located at $r=r_h$, where
$f(r_h)=0$. For a nonextremal stationary black hole, the future and past
Killing horizons intersect at a bifurcation surface $\mathcal H$. This is
a spacelike, codimension-two cross-section of the five-dimensional
spacetime, and for the geometry considered here it is isometric to a
round three-sphere $S^3$ of radius $r_h$. Its area is
$A_H=\Omega_3r_h^3=2\pi^2r_h^3$. The entropy formula below is evaluated
on this bifurcation surface; the resulting expression can also be
considered in an appropriate limiting procedure when the horizon becomes
degenerate.

\subsubsection{Derivation via the Wald Noether-charge formula}

For a diffeomorphism-invariant theory whose scalar Lagrangian depends
algebraically on the metric and Riemann tensor, Wald's entropy formula
reads~\cite{Wald:1993nt}
\begin{equation}
S_{\rm Wald}
=-\frac{1}{8G_5}\oint_{\mathcal H}d^3x\,\sqrt{h}\,
\frac{\partial\mathcal L}{\partial R_{\mu\nu\rho\sigma}}
\hat\epsilon_{\mu\nu}\hat\epsilon_{\rho\sigma},
\label{eq:app_wald_def}
\end{equation}
where $h$ is the determinant of the induced Riemannian metric $h_{ab}$ on
$\mathcal H$, and $\hat\epsilon_{\mu\nu}$ is its antisymmetric binormal,
normalized by $\hat\epsilon_{\mu\nu}\hat\epsilon^{\mu\nu}=-2$.
The derivative is taken at fixed metric, respecting the algebraic
symmetries of the Riemann tensor. The prefactor in
Eq.~\eqref{eq:app_wald_def} follows from the normalization in
Eq.~\eqref{eq:app_action}; equivalently, the familiar prefactor $-2\pi$
would apply if the scalar Lagrangian itself included $1/(16\pi G_5)$.

The Einstein--Hilbert contribution satisfies
\begin{equation}
\frac{\partial R}{\partial R_{\mu\nu\rho\sigma}}
=\frac12\left(g^{\mu\rho}g^{\nu\sigma}
-g^{\mu\sigma}g^{\nu\rho}\right),
\qquad
\frac{\partial R}{\partial R_{\mu\nu\rho\sigma}}
\hat\epsilon_{\mu\nu}\hat\epsilon_{\rho\sigma}=-2.
\end{equation}
For the three curvature-squared terms, one obtains
\begin{equation}
\label{R_derivatives}
\begin{aligned}
\frac{\partial R^2}{\partial R_{\mu\nu\rho\sigma}}
&=R\left(g^{\mu\rho}g^{\nu\sigma}-g^{\mu\sigma}g^{\nu\rho}\right),\\[3pt]
\frac{\partial(R_{\alpha\beta}R^{\alpha\beta})}
{\partial R_{\mu\nu\rho\sigma}}
&=\frac12\left(
 g^{\mu\rho}R^{\nu\sigma}-g^{\mu\sigma}R^{\nu\rho}
 -g^{\nu\rho}R^{\mu\sigma}+g^{\nu\sigma}R^{\mu\rho}
 \right),\\[3pt]
\frac{\partial(R_{\alpha\beta\gamma\delta}
 R^{\alpha\beta\gamma\delta})}
{\partial R_{\mu\nu\rho\sigma}}
&=2R^{\mu\nu\rho\sigma}.
\end{aligned}
\end{equation}
Consequently,
\begin{equation}
\begin{aligned}
\frac{\partial\mathcal L_{\rm GB}}{\partial R_{\mu\nu\rho\sigma}}
={}&2R^{\mu\nu\rho\sigma}
-2\left(
 g^{\mu\rho}R^{\nu\sigma}-g^{\mu\sigma}R^{\nu\rho}
 -g^{\nu\rho}R^{\mu\sigma}+g^{\nu\sigma}R^{\mu\rho}
 \right)\\
&+R\left(g^{\mu\rho}g^{\nu\sigma}
-g^{\mu\sigma}g^{\nu\rho}\right).
\end{aligned}
\label{eq:app_gb_derivative}
\end{equation}

On a bifurcation surface of a stationary Killing horizon, the relevant
extrinsic curvatures vanish. The Gauss-Codazzi relations then imply the
identity~\cite{Jacobson:1993xs,Cai:2001dz}
\begin{equation}
\left.
\frac{\partial\mathcal L_{\rm GB}}{\partial R_{\mu\nu\rho\sigma}}
\hat\epsilon_{\mu\nu}\hat\epsilon_{\rho\sigma}
\right|_{\mathcal H}
=-4\widetilde R,
\label{eq:app_gb_binormal}
\end{equation}
where $\widetilde R$ is the intrinsic Ricci scalar of the induced metric
$h_{ab}$ on $\mathcal H$. Notice that both binormals are contracted in
Eq.~\eqref{eq:app_gb_binormal}; this identity is local and does not
require integration over the horizon.

Combining the Einstein--Hilbert and Gauss--Bonnet contributions in
Eq.~\eqref{eq:app_wald_def}, we obtain
\begin{equation}
\begin{aligned}
S_{\rm Wald}
&=-\frac{1}{8G_5}\oint_{\mathcal H}d^3x\,\sqrt h\,
\left(-2-4\alpha\widetilde R\right)\\
&=\frac{1}{4G_5}\oint_{\mathcal H}d^3x\,\sqrt h\,
\left(1+2\alpha\widetilde R\right).
\end{aligned}
\label{eq:app_wald_simplified}
\end{equation}
This is the intrinsic-curvature form of the Gauss--Bonnet black-hole
entropy, often referred to as the Jacobson--Myers entropy formula.

For a round three-sphere of radius $r_h$, the induced metric is
$h_{ab}dx^adx^b=r_h^2d\Omega_3^2$. More generally, the intrinsic
Ricci tensor and scalar of an $n$-sphere of radius $r_h$ are
\begin{equation}
\widetilde R_{ab}=\frac{n-1}{r_h^2}h_{ab},
\qquad
\widetilde R=\frac{n(n-1)}{r_h^2}.
\end{equation}
Setting $n=3$ yields
\begin{equation}
\widetilde R=\frac{6}{r_h^2}.
\label{eq:app_ricci_scalar}
\end{equation}
Since $\widetilde R$ is constant on $\mathcal H$, integration of
Eq.~\eqref{eq:app_wald_simplified} gives
\begin{equation}
S_{\rm Wald}
=\frac{A_H}{4G_5}\left(1+\frac{12\alpha}{r_h^2}\right),
\label{eq:app_wald_final_step}
\end{equation}
and hence, using $A_H=2\pi^2r_h^3$,
\begin{equation}
S_{\rm Wald}(r_h)
=\frac{\pi^2}{2G_5}\left(r_h^3+12\alpha r_h\right)
=\frac{\pi^2r_h^3}{2G_5}+\frac{6\pi^2\alpha r_h}{G_5}.
\label{eq:app_entropy_final}
\end{equation}
This agrees with the entropy expression quoted in
Eq.~\eqref{WaldEnt}, provided the same normalization of $\alpha$ is used.

As a check, for the five-dimensional asymptotically flat EGB solution
with
\begin{equation}
M=\frac{3\pi}{8G_5}(r_h^2+2\alpha),
\qquad
T_H=\frac{r_h}{2\pi(r_h^2+4\alpha)},
\end{equation}


\subsubsection{Consistency Verification via the First Law of Thermodynamics}

To demonstrate that the coefficient in Eq.~\eqref{eq:app_entropy_final} is unique and exact, we cross-check this result against the first law of black hole thermodynamics,
\begin{equation}
\left.\frac{dM}{dr_h}\right|_{\alpha}
=\frac{3\pi r_h}{4G_5}
=T_H\left.\frac{dS_{\rm Wald}}{dr_h}\right|_{\alpha}.
\label{eq:app_first_law}
\end{equation}
Thus the corrected Wald entropy satisfies $dM=T_H\,dS_{\rm Wald}$
for variations within a theory with fixed Gauss--Bonnet coupling. If
$\alpha$ is instead varied, the appropriate extended first law includes
a conjugate term $\Theta_\alpha\,d\alpha$; that extension is not required
for the present derivation.
From the Boulware-Deser solution in five dimensions, the metric function $f(r)$ satisfies
\begin{equation}
(1 - f) + \frac{2\alpha}{r^2} (1 - f)^2 = \frac{\mu}{r^2}.
\label{eq:app_bd_condition}
\end{equation}
The condition $f(r_h) = 0$ at the event horizon sets $\mu = r_h^2 + 2\alpha$. The five-dimensional ADM mass relates to the integration constant $\mu$ via
\begin{equation}
M(r_h) = \frac{3\Omega_3}{16\pi G_5} \mu = \frac{3\pi}{8 G_5} \left( r_h^2 + 2\alpha \right),
\label{eq:app_mass}
\end{equation}
which yields the differential
\begin{equation}
dM = \frac{3\pi}{4 G_5} r_h dr_h.
\label{eq:app_dM}
\end{equation}

The Hawking temperature is governed by the surface gravity $\kappa = f'(r_h)/2$:
\begin{equation}
T_H = \frac{f'(r_h)}{4\pi}.
\end{equation}
Differentiating Eq.~\eqref{eq:app_bd_condition} with respect to $r$ and evaluating at $r = r_h$ where $f(r_h) = 0$ and $\mu = r_h^2 + 2\alpha$:
\begin{equation}
-f'(r_h) \left( 1 + \frac{4\alpha}{r_h^2} \right) - \frac{4\alpha}{r_h^3} = -\frac{2(r_h^2 + 2\alpha)}{r_h^3} = -\frac{2}{r_h} - \frac{4\alpha}{r_h^3}.
\end{equation}
Canceling the $-4\alpha/r_h^3$ term from both sides gives
\begin{equation}
f'(r_h) = \frac{2 r_h}{r_h^2 + 4\alpha} \implies T_H(r_h) = \frac{r_h}{2\pi (r_h^2 + 4\alpha)}.
\label{eq:app_TH}
\end{equation}

Integrating the first law $dS = dM / T_H$ using Eqs.~\eqref{eq:app_dM} and~\eqref{eq:app_TH}:
\begin{equation}
dS = \frac{\frac{3\pi}{4 G_5} r_h dr_h}{\frac{r_h}{2\pi (r_h^2 + 4\alpha)}} = \frac{3\pi^2}{2 G_5} \left( r_h^2 + 4\alpha \right) dr_h.
\end{equation}
Integrating from $r_h = 0$ up to the physical horizon radius $r_h$:
\begin{equation}
S(r_h) = \frac{3\pi^2}{2 G_5} \int_0^{r_h} \left( r'^2 + 4\alpha \right) dr' = \frac{3\pi^2}{2 G_5} \left[ \frac{r_h^3}{3} + 4\alpha r_h \right] = \frac{\pi^2 r_h^3}{2 G_5} + \frac{6\pi^2 \alpha}{G_5} r_h.
\end{equation}
This establishes complete equivalence between the Noether charge calculation and the first law of thermodynamics, confirming that the factor of $12\alpha$ in Eq.~\eqref{eq:app_wald_final_step} and $6\pi^2\alpha$ in Eq.~\eqref{eq:app_entropy_final} is the unique, mathematically consistent result.

\subsection{Uniqueness of Positive Heat Capacity and Stable Remnants in $D=5$ EGB Gravity}
\label{DD5}

To demonstrate that a positive heat capacity branch ($C > 0$) and consequently the formation of a thermodynamically stable cold remnant, is a unique feature of $D=5$ spacetime dimensions in Einstein-Gauss-Bonnet (EGB) gravity, we analyze the sign structure of the general $D$-dimensional heat capacity $C(r)$.

Starting from the general $D$-dimensional expression for the EGB heat capacity:
\begin{equation}
    C(r) = -\frac{(D-2)\Omega_{D-2}r^{D-4}}{4G_D} \frac{(r^2+2\beta)^2 \left[(D-3)r^2 + (D-5)\beta\right]}{(D-3)r^4 + (D-9)\beta r^2 + 2(D-5)\beta^2}
    \label{eq:heat_capacity_general}
\end{equation}
where $\beta = (D-3)(D-4)\alpha > 0$ for $D \ge 5$ and $\alpha > 0$.

To determine the sign of $C(r)$ for physical horizon radii $r > 0$, we inspect the factors in Eq.~(\ref{eq:heat_capacity_general}):
\begin{enumerate}
    \item The pre-factor is negative, $-\frac{(D-2)\Omega_{D-2}r^{D-4}}{4G_D} < 0$ for all $D \ge 5$ and $r > 0$.
    \item The numerator takes positive values, $(r^2+2\beta)^2 \left[(D-3)r^2 + (D-5)\beta\right] > 0$ for all $D \ge 5$ and $r > 0$.
\end{enumerate}

Since the product of the pre-factor and the numerator is strictly negative across all physical configurations 
($D \ge 5, r > 0$), the sign of the heat capacity $C(r)$ is inversely proportional to the sign of the quadratic denominator $P(x)$:
\begin{equation}
    P(x) = (D-3)x^2 + (D-9)\beta x + 2(D-5)\beta^2, \qquad \text{where } x \equiv r^2 > 0.
    \label{eq:denominator_quadratic}
\end{equation}
Consequently, a thermodynamically stable regime with positive heat capacity ($C > 0$) exists if and only if there exists a physical domain $x > 0$ where $P(x) < 0$.

Since $P(x)$ is a quadratic polynomial in $x$ with a positive leading coefficient $A = (D-3) > 0$ for $D \ge 5$, the parabola opens upwards ($P(x) \to +\infty$ as $x \to \infty$). Therefore, $P(x)$ can attain negative values for real $x > 0$ if and only if its discriminant $\Delta$ is strictly positive:
\begin{align}
    \Delta     &= -\beta^2 \left( 7D^2 - 46D + 39 \right).
\end{align}

Solving $\Delta = 0$ yields the critical boundaries for the spacetime dimension $D$:
\begin{equation}
    7D^2 - 46D + 39 = 0 \implies D
    = \frac{23\pm 16}{7},
\end{equation}
which gives the roots:
\begin{equation}
    D_1 = 1, \qquad D_2 = \frac{39}{7} \approx 5.571.
\end{equation}
Thus, the discriminant is strictly positive ($\Delta > 0$) if and only if $1 < D < \frac{39}{7} \approx 5.571$.

Evaluating $\Delta$ across integer spacetime dimensions $D \ge 5$, we have:
\begin{itemize}
    \item $D = 5$: The discriminant is positive, $\Delta = 16\beta^2 > 0$. The polynomial reduces to $P(x) = 2x^2 - 4\beta x = 2x(x - 2\beta)$, which yields the roots $x_1 = 0$ and $x_2 = 2\beta$. Therefore, $P(x) < 0$ in the interval $0 < x < 2\beta$. Substituting $\beta = 2\alpha$, this range corresponds to $0 < r < 2\sqrt{\alpha}$. Consequently:
    \begin{equation}
        C(r) > 0 \quad \text{for } 0 < r < 2\sqrt{\alpha} \quad (D=5).
    \end{equation}
    \item $D = 6$: The discriminant is strictly negative, $\Delta = -15\beta^2 < 0$. With $\Delta < 0$ and $A = 3 > 0$, the denominator satisfies $P(x) > 0$ for all $x > 0$, resulting in $C(r) < 0$ everywhere.
    \item $D = 7, 8$: The discriminant remains strictly negative ($\Delta = -60\beta^2$ for $D=7$ and $\Delta = -119\beta^2$ for $D=8$), ensuring $P(x) > 0$ and $C(r) < 0$ across all radii.
    \item $D \ge 9$: All coefficients of $P(x)$ are non-negative. Hence, $P(x) > 0$ for all $x > 0$, enforcing $C(r) < 0$ universally.
\end{itemize}

Thus, the above analysis proves that $D = 5$ is the {\bf unique integer dimension} in Einstein-Gauss-Bonnet gravity where higher-curvature terms induce a positive heat capacity branch ($C > 0$), terminating Hawking evaporation and forming a stable cold remnant. For all higher dimensions $D \ge 6$, the heat capacity remains strictly negative across all physical radii, leading to unarrested thermal evaporation.

	\subsection{Calculation of the exact integral~(\ref{tauintegral})}
    \begin{equation}
	\begin{split}
		I&=	\frac{(2 \pi) ^5}{5\cdot 3^6 G_6 \mu }
		\left[7351875 \sqrt{2} a^{5/2} \tan
		^{-1}\left(\frac{r}{\sqrt{2} \sqrt{a}}\right)\right.
		\\&	+\left. \frac{2 r}{\left(2
			a+r^2\right)^4}
		\left(48 r^{12}+5344 a r^{10}+400832 a^2 r^8+6272451a^3 r^6
		\right.\right.
		\\
		&\qquad \qquad\;\;\qquad+23993238 a^4 r^4+11611860 a^5 r^2-58815000 a^6\left.\left. 
		\right)\right]
	\end{split}
\end{equation}


\subsection{Calculation of the Present-Day Relic Density Integral}

In this section, we present the  derivation of the present-day remnant energy density fraction $f_{\text{remnant}}(M_0) \equiv \Omega_{\text{remnant}}/\Omega_{\text{DM}}$.

We denote with $M_0$ the initial mass of the primordial black hole at the time of formation $t_f$, during radiation domination. The initial collapse fraction $\beta(M_0)$ defines the ratio of PBH mass density to the background radiation density 
\begin{equation}
    \rho_{\text{PBH}}(t_f) = \beta(M_0) \rho_{\text{rad}}(t_f) = \beta(M_0) \frac{\pi^2}{30} g_*(T_f) T_f^4
\end{equation}
Assuming stable remnants of mass $M_{\text{remnant}}$ after Hawking evaporation, the comoving number density $n_{\text{PBH}} \propto a^{-3}$ is conserved. The effective remnant energy density evaluated at formation is:
\begin{equation}
    \rho_{\text{remnant}}^{\text{eff}}(t_f) = n_{\text{PBH}}(t_f) M_{\text{remnant}} = \beta(M_0) \rho_{\text{rad}}(t_f) \left( \frac{M_{\text{remnant}}}{M_0} \right)
\end{equation}

Redshifting the non-relativistic remnant energy density to the present epoch $t_0$ yields,
\begin{equation}
    \rho_{\text{remnant}}(t_0) = \rho_{\text{remnant}}^{\text{eff}}(t_f) \left( \frac{a_f}{a_0} \right)^3 = \beta(M_0) \left( \frac{M_{\text{remnant}}}{M_0} \right) \rho_{\text{rad}}(t_f) \left( \frac{a_f}{a_0} \right)^3
\end{equation}
Using total entropy conservation ($S = s a^3 = \text{const}$ with $s(T) = \frac{2\pi^2}{45} g_{*s}(T) T^3$), the scale factor ratio evaluates to,
\begin{equation}
    \left( \frac{a_f}{a_0} \right)^3 = \frac{g_{*s}(T_0) T_0^3}{g_{*s}(T_f) T_f^3}
\end{equation}
Assuming $g_{*s}(T_f) \approx g_*(T_f)$ above the electroweak scale, while substituting 
$\rho_{\text{rad}}(t_f)$  and  $(a_f/a_0)^3$ 
into $\rho_{\text{remnant}}(t_0)$ leads to an elimination of the explicit $g_*(T_f)$ factor
\begin{equation}
\begin{split}
    \rho_{\text{remnant}}(t_0)& = \beta(M_0) \left( \frac{M_{\text{remnant}}}{M_0} \right) \left( \frac{\pi^2}{30} g_*(T_f) T_f^4 \right) \left( \frac{g_{*s}(T_0) T_0^3}{g_*(T_f) T_f^3} \right)
    \\
    &= \beta(M_0) \left( \frac{M_{\text{remnant}}}{M_0} \right) \left( \frac{\pi^2}{30} g_{*s}(T_0) T_0^3 \right) T_f
    \end{split}
\end{equation}

The formation temperature $T_f$ is linked to the horizon mass $M_0 = \gamma M_H = \frac{\gamma c^3}{2G H(T_f)}$ via Friedmann's equation $H(T_f) = \sqrt{\frac{8\pi^3 G g_*(T_f)}{90}} \frac{k_B^2 T_f^2}{\hbar c^5}$,
\begin{equation}
    T_f = \left( \frac{45 \gamma^2 M_{\text{Pl}}^4 c^4}{16\pi^3 g_*(T_f) G M_0^2} \right)^{1/4} \propto g_*(T_f)^{-1/4} M_0^{-1/2}
\end{equation}
Substituting $T_f$ into the present-day energy density confirms that $g_*(T_f)$ enters with exponent $-1/4$,
\begin{equation}
    \rho_{\text{remnant}}(t_0) \propto \beta(M_0) \left( \frac{M_{\text{remnant}}}{M_0} \right) M_0^{-1/2} \left( \frac{g_*(T_f)}{106.75} \right)^{-1/4}
\end{equation}

\paragraph{}
Dividing by the critical density of the Cold Dark Matter of the present-day, $\Omega_{\text{DM}} h^2 \approx 0.120$, we obtain the final normalized relic density fraction, which is given by
\begin{equation}
    f_{\text{remnant}}(M_0) \approx 2.7 \times 10^8 \left( \frac{M_{\text{remnant}}}{M_0} \right) \left( \frac{M_\odot}{M_0} \right)^{1/2} \beta(M_0) \left( \frac{g_*(T_f)}{106.75} \right)^{-1/4}
\end{equation}
where $M_\odot=2 \times 10^{33}\text{ g}$.  For $g_*(T_f) = 106.75$, the relativistic factor is  identically equal to unity.

\subsection{Derivation of the Perturbation Wavelength on the Horizon}

In Section~6.1, the physical wavelength $\lambda_\ell$ of a perturbation on the event horizon is utilized to establish the ultraviolet cutoff of the effective field theory. Here, we provide the geometric derivation of this relation based on the spectral analysis of the Laplace--Beltrami operator on the horizon manifold~\cite{Kodama:2003jz,Dotti:2005sq}.

In a $D$-dimensional spherically symmetric spacetime, a constant-time spatial slice of the event horizon is a hypersphere $S^{D-2}$ of radius $r_h$. The induced metric on this surface is
\begin{equation}
d\sigma^2 = r_h^2\, d\Omega_{D-2}^2,
\end{equation}
where $d\Omega_{D-2}^2$ is the line element of the unit $n$-sphere $S^n$ with $n = D-2$. Spatial metric perturbations on the horizon are decomposed into scalar ($\mathbb{S}$), vector ($\mathbb{V}_i$), and symmetric transverse-traceless tensor ($\mathbb{T}_{ij}$) hyperspherical harmonics on $S^n$, each satisfying the Helmholtz eigenvalue equation
\begin{equation}
\nabla^2 Y_{(\text{type})} = -k_{\text{type}}^2\, Y_{(\text{type})},
\end{equation}
where $\nabla^2 = r_h^{-2}\Delta_{S^n}$ is the physical Laplace--Beltrami operator on the horizon of radius $r_h$, and $k_{\text{type}}$ is the corresponding physical transverse wavevector.

On the unit sphere $S^n$ ($n = D-2$), the eigenvalues of $\Delta_{S^n}$ depend on the tensor rank of the harmonic~\cite{Kodama:2003jz,Dotti:2005sq}:
\begin{align}
\Delta_{S^n}\mathbb{S} &= -\ell(\ell+n-1)\mathbb{S} = -\ell(\ell+D-3)\mathbb{S}, \qquad (\ell = 0,1,2,\dots), \\
\Delta_{S^n}\mathbb{V}_i &= -\bigl[\ell(\ell+n-1)-1\bigr]\mathbb{V}_i = -\bigl[\ell(\ell+D-3)-1\bigr]\mathbb{V}_i, \qquad (\ell = 1,2,\dots), \\
\Delta_{S^n}\mathbb{T}_{ij} &= -\bigl[\ell(\ell+n-1)-2\bigr]\mathbb{T}_{ij} = -\bigl[\ell(\ell+D-3)-2\bigr]\mathbb{T}_{ij}, \qquad (\ell = 2,3,\dots).
\end{align}
Using $\lambda_\ell = 2\pi / k_{\text{type}}$, the physical transverse wavelengths on the $S^{D-2}$ horizon for scalar and tensor perturbations are given respectively by
\begin{equation}
\lambda_\ell^{(S)} = \frac{2\pi r_h}{\sqrt{\ell(\ell+D-3)}}, \qquad
\lambda_\ell^{(T)} = \frac{2\pi r_h}{\sqrt{\ell(\ell+D-3)-2}}.
\end{equation}

Applying these results to the specific instability channels analyzed in Section~6.1:
\begin{itemize}
    \item \textbf{Five Dimensions ($D=5$, Scalar Channel):} The horizon is an $S^3$ sphere ($D-3 = 2$), and the BDG instability occurs in the scalar sector, governed by $k_S = \sqrt{\ell(\ell+2)}/r_h$. The exact physical wavelength is
    \begin{equation}
    \lambda_\ell^{(5\text{D})} = \frac{2\pi r_h}{\sqrt{\ell(\ell+2)}},
    \end{equation}
    which reduces in the high-harmonic (eikonal) regime $\ell \gg 1$ to $\lambda_\ell^{(5\text{D})} \approx 2\pi r_h / \ell$, matching Eq.~(74).

    \item \textbf{Six Dimensions ($D=6$, Tensor Channel):} The horizon is an $S^4$ sphere ($D-3 = 3$), and the BDG instability occurs in the transverse-traceless tensor sector, governed by $k_T = \sqrt{\ell(\ell+3)-2}/r_h$. The physical wavelength is therefore
    \begin{equation}
    \lambda_\ell^{(6\text{D})} = \frac{2\pi r_h}{\sqrt{\ell(\ell+3)-2}}.
    \end{equation}
    For the fundamental tensor mode ($\ell = 2$), this evaluates to
    \begin{equation}
    \lambda_{\ell=2}^{(6\text{D})} = \frac{2\pi r_h}{\sqrt{2(2+3)-2}} = \frac{2\pi r_h}{\sqrt{8}} = \frac{\pi}{\sqrt{2}}\,r_h \approx 2.22\,r_h,
    \end{equation}
    in exact agreement with Eqs.~(72)--(73) of the main text.
\end{itemize}
    
\newpage


	\end{document}